\documentclass[twocolumn,american,aps,prb,superscriptaddress]{revtex4-2}
\usepackage[T1]{fontenc}
\usepackage[utf8]{inputenc}
\usepackage{units}
\usepackage{amsmath}
\usepackage{amssymb}
\usepackage{graphicx}

\makeatletter
\usepackage{babel}

\makeatother

\usepackage{babel}
\begin{document}
\title{Magneto-optical Kerr effect in pump-probe setups}
\author{Amir Eskandari-asl}
\affiliation{Dipartimento di Fisica ``E.R. Caianiello'', Università degli Studi
di Salerno, I-84084 Fisciano (SA), Italy}
\author{Adolfo Avella}
\affiliation{Dipartimento di Fisica ``E.R. Caianiello'', Università degli Studi
di Salerno, I-84084 Fisciano (SA), Italy}
\affiliation{CNR-SPIN, Unità di Salerno, I-84084 Fisciano (SA), Italy}
\affiliation{CNISM, Unità di Salerno, Università degli Studi di Salerno, I-84084
Fisciano (SA), Italy}
\begin{abstract}
We develop a general theoretical framework for computing the time-resolved
magneto-optical Kerr effect in ultrafast pump--probe setups, formulated
within the Dynamical Projective Operatorial Approach (DPOA) and its
application to the generalized linear-response theory for pumped systems.
Furthermore, we exploit this formalism to express the post-pump optical
conductivity --- and consequently the Kerr rotation --- in terms
of the time-evolved single-particle density matrix (SPDM), providing
a transparent and computationally efficient description of photo-excited
multi-band systems. This extension, in addition to its lower computational
cost, has the advantage of allowing the inclusion of phenomenological
damping. We illustrate the formalism using both (i) a two-band tight-binding
model, which captures the essential physics of ultrafast spin--charge
dynamics and the Kerr rotation, and (ii) weakly spin-polarized germanium,
as a realistic playground with a complex band structure. The results
demonstrate that, by exploiting DPOA and/or its SPDM extension, one
can reliably reproduce both the short-time features under the pump-pulse
envelope and the long-time dynamics after excitation, offering a versatile
framework for analyzing time-resolved magneto-optical Kerr effect
experiments in complex materials. Moreover, this analysis clearly
shows that the Kerr rotation can be used to deduce experimentally
the relevant n-photon resonances for a given specific material.
\end{abstract}
\maketitle

\section{Introduction}

In recent years, the development of ultrafast pump--probe spectroscopy
has opened a new frontier in exploring the dynamics of condensed matter
systems on femtosecond and even sub-femtosecond time scales \citep{brabec2000intense,krausz2009attosecond,krausz2014attosecond,calegari2016advances,gandolfi2017emergent,borrego2022attosecond,inzani2023field,inzani2023photoinduced}.
By monitoring the temporal evolution of photo-excited carriers, these
techniques provide direct access to the microscopic mechanisms governing
electronic, spin, and lattice degrees of freedom out of equilibrium.
From a technological standpoint, such insight is essential for advancing
next-generation ultrafast optoelectronic and spintronic devices. From
a fundamental perspective, it enables one to probe pulse symmetry
breaking, coherence, and relaxation processes in real time \citep{Zurch_17,PhysRevB.97.205202,perfetti2008femtosecond}.

Among various ultrafast observables, the time-resolved magneto-optical
Kerr effect has emerged as a powerful probe for detecting transient
magnetization dynamics and non-equilibrium spin polarization \citep{ZhangHubnerLefkidisBaiGeorge2009_TRMOKE}.
The Kerr and also Faraday rotations---arising from the off-diagonal
components of the optical conductivity tensor---encode information
about spin--orbit coupling (SOC) and the instantaneous breaking of
time-reversal symmetry (TRS) \citep{sato2022fundamentals}. They thus
provide a direct experimental window into ultrafast processes such
as light-induced magnetization, coherent spin precession, and photo-induced
topological transitions \citep{beaurepaire1996ultrafast,kimel2005ultrafast,wang2006ultrafast,kirilyuk2010ultrafast,WuJiangWeng2010_SpinDynamics,hennecke2022ultrafast}.
In systems such as pumped altermagnets or magnetic semiconductors,
Kerr rotation measurements serve as a sensitive fingerprint of the
evolving symmetry and band topology on ultrafast time scales \citep{nemec2018antiferromagnetic,gray2024time,eskandari2025controlling}.

Despite its experimental relevance, a comprehensive and computationally
affordable theoretical framework for evaluating the transient Kerr
response has remained elusive \citep{kampfrath2013terahertz,zhang2020nonequilibrium}.
Ab initio methods based on time-dependent density functional theory
are, in principle, capable of describing the pump--probe dynamics
\citep{de2013inside,de2016monitoring,de2018real}, but their extremely
high computational cost and limited interpretability make them unsuitable
for exploring a broad parameter space or for isolating the microscopic
origin of specific phenomena. Model-Hamiltonian approaches \citep{schlaepfer2018attosecond,sato2018role,broers2022detecting}
can address these issues by allowing controlled manipulation of individual
Hamiltonian terms, but they often lack the generality and the complexity
required for realistic simulations.

To bridge this gap, we recently developed the Dynamical Projective
Operatorial Approach (DPOA) \citep{inzani2023field,eskandari2024time,eskandari2024generalized,eskandari2025dynamical,eskandari2024out},
an efficient operator-based formalism for studying ultrafast phenomena
in realistic multi-band systems. DPOA enables the real-time evolution
of composite operators under the action of intense pump fields, while
maintaining full access to microscopic observables such as the single-particle
density matrix (SPDM), band populations, inter-band coherences, and
in principle, every multi-time multi-particle response function, including
time-resolved angle-resolved photo-emission spectroscopy \citep{eskandari2024time}
and transient optical properties, through the generalized linear response
theory for pumped systems \citep{eskandari2024generalized}. DPOA
formulation is in principle, fully general---applicable to systems
with arbitrary lattice structures, numbers of bands, or other complexities.

In the present work, we build upon DPOA to formulate a general theory
for computing the transient Kerr rotation in pump--probe setups.
Furthermore, we extend the framework to include phenomenological damping
within the SPDM dynamics, allowing us to capture relaxation effects
that are essential in realistic experimental conditions.

The paper is organized as follows. In Sec.~\ref{sec:Theory}, we
outline the theoretical formulation of the two-time optical conductivity
and its reduction in terms of the SPDM. In Sec.~\ref{sec:Numerical-studies},
we apply the formalism to a two-band model to demonstrate its simulation
capability and to analyze the transient optical conductivity and Kerr
rotation under different pump pulse conditions. In Sec.~\ref{sec:germanium},
we apply the formalism to a weakly spin-polarized germanium to demonstrate
the capability of the formalism to tackle the complexity of a real
material and of a realistic band structure. Finally, Sec.~\ref{sec:Conclusions}
summarizes our findings and discusses possible extensions of the present
framework to more complex materials and correlated systems.

\section{Theory}\label{sec:Theory}

\subsection{The pumped system}

Although this framework can be applied also to single atoms, molecules,
photonic lattices, and many other kind of systems, in this manuscript,
we focus on a lattice system driven by a linearly polarized electromagnetic
pulse described by a vector potential $\boldsymbol{A}_{\text{pu}}\left(t\right)=A_{\text{pu}}\left(t\right)\hat{\boldsymbol{u}}_{\text{pu}}$
and the corresponding electric field $\boldsymbol{E}_{\text{pu}}\left(t\right)=E_{\text{pu}}\left(t\right)\hat{\boldsymbol{u}}_{\text{pu}}=-\partial_{t}A_{\text{pu}}\left(t\right)\hat{\boldsymbol{u}}_{\text{pu}}$,
where $\hat{\boldsymbol{u}}_{\text{pu}}$ identifies the polarization
direction of the pump pulse. The pump pulse is assumed to act on the
system starting from an initial time $t_{\mathrm{ini}}\rightarrow-\infty$.
While we focus on linear pump polarization for simplicity in the following
discussion, the formalism can be readily extended to account for a
circularly polarized pump pulse.

In the dipole gauge, the time-dependent Hamiltonian of such a system
takes the form \citep{schuler2021gauge,eskandari2024time,eskandari2024generalized}
\begin{equation}
\mathcal{H}\left(t\right)=\sum_{\mathbf{k},n,n{}^{\prime}}c_{\mathbf{k},n}^{\dagger}\left(t\right)\Xi_{\mathbf{k},n,n{}^{\prime}}\left(t\right)c_{\mathbf{k},n{}^{\prime}}\left(t\right),
\end{equation}
where $c_{\mathbf{k},n}\left(t\right)$ annihilates an electron with
crystal momentum $\mathbf{k}$ and band index $n$; in the following,
the index $n$ will implicitly include the spin degree of freedom.
The matrix $\Xi_{\mathbf{k}}\left(t\right)$ is obtained by the first-quantization
single-particle Hamiltonian and reads as 
\begin{equation}
\Xi_{\mathbf{k}}\left(t\right)=T_{\boldsymbol{k}}\left(t\right)+\mathrm{e}E_{\text{pu}}\left(t\right)D_{\boldsymbol{k},\text{pu}}\left(t\right),\label{eq:ham_dg-b}
\end{equation}
where $D_{\boldsymbol{k},\text{pu}}\left(t\right)=\boldsymbol{D}_{\boldsymbol{k}}\left(t\right)\cdot\hat{\boldsymbol{u}}_{\text{pu}}$.
Here, $T_{\boldsymbol{k}}\left(t\right)$ and $\boldsymbol{D}_{\boldsymbol{k}}\left(t\right)$
denote the time-dependent hopping and dipole matrices, respectively.
They are constructed by applying the Peierls substitution in a localized
basis (usually, the Wannier basis) and then transforming to the band
basis in which the equilibrium Hamiltonian is diagonal \citep{eskandari2024time,eskandari2024generalized}.
Hereafter, we distinguish operators written in the localized basis
by an over-script $\sim$.

The transformation from the localized to the band basis is implemented
via the unitary matrix $\Omega_{\mathbf{k}}$. In compact matrix notation,
$\Omega_{\mathbf{k}}^{\dagger}\cdot\tilde{T}_{\mathbf{k}}\cdot\Omega_{\mathbf{k}}$
is diagonal, with the band energies $\varepsilon_{\mathbf{k},n}$
appearing on the diagonal. For a generic matrix $M$ (representing
any of $T$, $\boldsymbol{D}$, or their derivatives with respect
to $\mathbf{k}$), we write
\begin{equation}
M_{\mathbf{k}}\left(t\right)=\Omega_{\mathbf{k}}^{\dagger}\cdot\tilde{M}_{\mathbf{k}+\frac{\mathrm{e}}{\hbar}\boldsymbol{A}_{\text{pu}}\left(t\right)}\cdot\Omega_{\mathbf{k}}.
\end{equation}
The symbol $\cdot$ is used to denote both vector products in Cartesian
space and matrix products in the electronic Hilbert space. In realistic
multi-band systems, evaluated over dense $\mathbf{k}$-grids, it is
computationally advantageous to expand $M_{\mathbf{k}+\frac{\mathrm{e}}{\hbar}\boldsymbol{A}_{\text{pu}}\left(t\right)}$
in powers of the vector potential, by the so-called Peierls expansion
\citep{eskandari2024time}.

Under the application of the pump pulse, DPOA allows us to express
the time-evolved operators $c_{\mathbf{k}}\left(t\right)$ in the
Heisenberg picture in terms of their equilibrium counterparts $c_{\mathbf{k}}\left(t_{\mathrm{ini}}\right)$
(where $c_{\mathbf{k}}\left(t\right)$ denotes a column vector whose
components are the operators $c_{\mathbf{k},n}\left(t\right)$). This
is achieved through the projection matrices $P_{\mathbf{k}}\left(t\right)$,
such that
\begin{equation}
c_{\mathbf{k}}\left(t\right)=P_{\mathbf{k}}\left(t\right)\cdot c_{\mathbf{k}}\left(t_{\mathrm{ini}}\right).
\end{equation}
The projection matrices satisfy the following equation of motion:
\begin{equation}
\mathrm{i}\hbar\partial_{t}P_{\mathbf{k}}\left(t\right)=\Xi_{\mathbf{k}}\left(t\right)\cdot P_{\mathbf{k}}\left(t\right),\label{eq:EOM_P}
\end{equation}
which is solved numerically with the initial condition $P_{\mathbf{k}}\left(t_{\mathrm{ini}}\right)=\boldsymbol{1}$.

\subsection{Transient optical conductivity}

Besides the pump pulse polarization, $\hat{\boldsymbol{u}}_{\text{pu}}$,
we also introduce the incoming probe pulse polarization, $\hat{\boldsymbol{u}}_{\text{in}}$,
and the outgoing probe pulse polarization, $\hat{\boldsymbol{u}}_{\text{out}}$.
We also introduce a compact notation for the projection of vectors
and rank 2 tensors in Cartesian space on unit vectors: $v_{\mathrm{a}}=\boldsymbol{v}\cdot\hat{\boldsymbol{u}}_{\mathrm{a}}$
and $s_{\mathrm{a},\mathrm{b}}=\hat{\boldsymbol{u}}_{\mathrm{a}}\cdot\boldsymbol{s}\cdot\hat{\boldsymbol{u}}_{\mathrm{b}}$,
where $\mathrm{a}$ and $\mathrm{b}$ denote polarization directions
such as $\mathrm{pu}$, $\mathrm{in}$, $\mathrm{out}$, $x$ and
$y$.

Within the generalized linear response theory \citep{eskandari2024generalized},
and by employing DPOA, the out-of-equilibrium two-time optical conductivity
of a pumped system probed at time $t_{\mathrm{pr}}$ (the center of
the probe pulse), $\sigma_{\text{out},\text{in}}\left(t,t_{\mathrm{pr}}\right)$,
can be written as \citep{eskandari2024generalized} 
\begin{equation}
\sigma_{\text{out},\text{in}}\left(t,t_{\mathrm{pr}}\right)=\sigma_{\text{out},\text{in}}^{\left(1\right)}\left(t,t_{\mathrm{pr}}\right)+\sigma_{\text{out},\text{in}}^{\left(2\right)}\left(t,t_{\mathrm{pr}}\right).\label{eq:sigma-P-t}
\end{equation}
The first contribution reads as
\begin{multline}
\sigma_{\text{out},\text{in}}^{\left(1\right)}\left(t,t_{\mathrm{pr}}\right)=\frac{\mathrm{i}e}{\hbar\mathcal{V}}\theta\left(t-t_{\mathrm{pr}}\right)\sum_{\boldsymbol{k}}\mathrm{Tr}\left\{ Z_{\boldsymbol{k}}\left(t\right)\cdot\right.\\
\left.\cdot\left[Y_{\boldsymbol{k}}\left(t\right)-Y_{\boldsymbol{k}}\left(t_{\mathrm{pr}}\right)+X_{\boldsymbol{k}}\left(t_{\mathrm{pr}}\right)\right]\right\} ,\label{eq:sig_ZYX}
\end{multline}
where $\theta\left(t-t_{\mathrm{pr}}\right)$ is the Heaviside theta
function, $N_{\mathrm{grid}}$ is the number of points sampling the
Brillouin zone (BZ), $e>0$ is the electronic charge, and $\mathcal{V}=N_{\mathrm{grid}}v_{\mathrm{uc}}$
is the total system volume, in which $v_{\mathrm{uc}}$ is the unit-cell
volume. The matrices $Z_{\boldsymbol{k}}\left(t\right)$, $Y_{\boldsymbol{k}}\left(t\right)$
and $X_{\boldsymbol{k}}\left(t\right)$ enable efficient numerical
calculations and are defined as \citep{eskandari2024generalized}
\begin{equation}
Z_{\boldsymbol{k}}\left(t\right)=\left(P_{\boldsymbol{k}}^{\dagger}\left(t\right)\cdot J_{\boldsymbol{k},\text{out}}\left(t\right)\cdot P_{\boldsymbol{k}}\left(t\right)\right)\circ\Delta F_{\boldsymbol{k}},
\end{equation}
\begin{equation}
Y_{\boldsymbol{k}}\left(t\right)=-\frac{e}{\hbar}\int_{t_{\mathrm{ini}}}^{t}dt^{\prime}P_{\boldsymbol{k}}^{\dagger}\left(t^{\prime}\right)\cdot V_{\boldsymbol{k},\text{in}}\left(t^{\prime}\right)\cdot P_{\boldsymbol{k}}\left(t^{\prime}\right),\label{eq:Y}
\end{equation}
\begin{equation}
X_{\boldsymbol{k}}\left(t\right)=eP_{\boldsymbol{k}}^{\dagger}\left(t\right)\cdot D_{\boldsymbol{k},\text{in}}\left(t\right)\cdot P_{\boldsymbol{k}}\left(t\right).\label{eq:X}
\end{equation}
Here $\varDelta F_{\boldsymbol{k},n,n^{\prime}}=f_{\boldsymbol{k},n}-f_{\boldsymbol{k},n^{\prime}}$,
with $f_{\boldsymbol{k},n}=\frac{1}{\mathrm{1+e^{-\beta\left(\varepsilon_{\mathbf{k},n}-\mu\right)}}}$
the Fermi function, which determines the initial band occupations
as the equilibrium thermal distribution, $\mu$ the chemical potential
and $\beta$ the inverse temperature. The symbol $\circ$ denotes
Hadamard (element-wise) matrix product. The current-related matrices
are
\begin{equation}
J_{\boldsymbol{k},\mathrm{a}}\left(t\right)=\frac{1}{\hbar}\eta_{\boldsymbol{k},\mathrm{a}}\left(t\right)-\frac{\mathrm{i}}{\hbar}\left[D_{\boldsymbol{k},\mathrm{a}}\left(t\right),T_{\boldsymbol{k}}\left(t\right)\right],
\end{equation}
\begin{equation}
\eta_{\boldsymbol{k},\mathrm{a}}\left(t\right)=\Omega_{\boldsymbol{k}}^{\dagger}\cdot\left(\hat{\boldsymbol{u}}_{\mathrm{a}}\cdot\boldsymbol{\nabla}_{\boldsymbol{k}}\tilde{T}_{\boldsymbol{k}+\frac{e}{\hbar}\boldsymbol{A}_{\mathrm{pu}}\left(t\right)}\right)\cdot\Omega_{\boldsymbol{k}},
\end{equation}
\begin{equation}
V_{\boldsymbol{k},\mathrm{a}}\left(t\right)=\eta_{\boldsymbol{k},\mathrm{a}}\left(t\right)+e\Lambda_{\boldsymbol{k},\mathrm{a},\text{pu}}\left(t\right)E_{\mathrm{pu}}\left(t\right),\label{eq:Vk}
\end{equation}
\begin{equation}
\Lambda_{\boldsymbol{k},\mathrm{a},\mathrm{b}}\left(t\right)=\Omega_{\boldsymbol{k}}^{\dagger}\cdot\left(\hat{\boldsymbol{u}}_{\mathrm{a}}\cdot\boldsymbol{\nabla}_{\boldsymbol{k}}\tilde{D}_{\boldsymbol{k}+\frac{e}{\hbar}\boldsymbol{A}_{\mathrm{pu}}\left(t\right),\mathrm{b}}\right)\cdot\Omega_{\boldsymbol{k}},\label{eq:VD}
\end{equation}
where $[\varPhi,\varPsi]$ denotes the commutator.

The second contribution reads as 
\begin{multline}
\sigma_{\text{out},\text{in}}^{\left(2\right)}\left(t,t_{\mathrm{pr}}\right)=\frac{e}{\mathcal{V}}\theta\left(t-t_{\mathrm{pr}}\right)\times\\
\times\sum_{\boldsymbol{k}}\text{Tr}\left\{ \left(\frac{\delta J_{\boldsymbol{k}}}{\delta A}\right)_{\text{out},\text{in}}\left(t\right)\cdot\rho_{\boldsymbol{k}}\left(t\right)\right\} ,\label{eq:sigma-2-P-t}
\end{multline}
where $\rho_{\boldsymbol{k}}\left(t\right)$ is the single-particle
density matrix (SPDM): 
\begin{equation}
\rho_{\boldsymbol{k},n,n^{\prime}}\left(t\right)=\left\langle c_{\boldsymbol{k},n^{\prime}}^{\dagger}\left(t\right)c_{\boldsymbol{k},n}\left(t\right)\right\rangle .
\end{equation}
It can be readily shown that \citep{eskandari2024time} 
\begin{equation}
\rho_{\boldsymbol{k}}\left(t\right)=P_{\boldsymbol{k}}\left(t\right)\cdot F_{\boldsymbol{k}}\cdot P_{\boldsymbol{k}}^{\dagger}\left(t\right),\label{eq:rho-P}
\end{equation}
where $F_{\boldsymbol{k},n,n^{\prime}}=\delta_{n,n^{\prime}}f_{\boldsymbol{k},n}$.
Furthermore, 
\begin{multline}
\left(\frac{\delta J_{\boldsymbol{k}}}{\delta A}\right)_{\text{out},\text{in}}\left(t\right)=\frac{e}{\hbar^{2}}\xi_{\boldsymbol{k},\text{out},\text{in}}\left(t\right)\\
-\frac{\mathrm{i}e}{\hbar^{2}}\left[\Lambda_{\boldsymbol{k},\text{in},\text{out}}\left(t\right),T_{\boldsymbol{k}}\left(t\right)\right]\\
-\frac{\mathrm{i}e}{\hbar^{2}}\left[D_{\boldsymbol{k},\text{out}}\left(t\right),\eta_{\boldsymbol{k},\text{in}}\left(t\right)\right],
\end{multline}
in which 
\begin{equation}
\xi_{\boldsymbol{k},\mathrm{a},\mathrm{b}}\left(t\right)=\Omega_{\boldsymbol{k}}^{\dagger}\cdot\left[\left(\hat{\boldsymbol{u}}_{\mathrm{a}}\cdot\boldsymbol{\nabla}_{\boldsymbol{k}}\right)\left(\hat{\boldsymbol{u}}_{\mathrm{b}}\cdot\boldsymbol{\nabla}_{\boldsymbol{k}}\right)\tilde{T}_{\boldsymbol{k}+\frac{e}{\hbar}\boldsymbol{A}_{\mathrm{pu}}\left(t\right)}\right]\cdot\Omega_{\boldsymbol{k}}.
\end{equation}

Once the full time-dependent optical conductivity, $\sigma_{\text{out},\text{in}}\left(t,t_{\mathrm{pr}}\right)$,
is obtained, it is often more convenient to work in the frequency
domain. This is done by performing a Fourier transform with respect
to the relative time $t-t_{\mathrm{pr}}$, thus expressing the conductivity
as a function of the probe pulse frequency $\omega$: 
\begin{equation}
\sigma_{\text{out},\text{in}}\left(\omega,t_{\mathrm{pr}}\right)=\int_{-\infty}^{+\infty}\mathrm{e}^{\mathrm{i}\left(\omega+\mathrm{i}\lambda\right)\left(t-t_{\mathrm{pr}}\right)}\sigma_{\text{out},\text{in}}\left(t,t_{\mathrm{pr}}\right)dt,\label{eq:FT-sigma}
\end{equation}
where $\lambda$ is a positive damping factor, which in some other
works is dubbed $0^{+}$. Since the optical conductivity is causal
and thus proportional to the Heaviside function, the lower limit of
the integral, $-\infty$, can be replaced by $t_{\mathrm{pr}}$.

A substantial simplification in the numerical evaluation of Eq.~\eqref{eq:FT-sigma}
is obtained by observing that the pump pulse extends over a finite
time interval. Specifically, one can restrict the upper limit of the
integral to a cutoff time $t_{\mathrm{fin}}>t_{\mathrm{pr}}$, chosen
such that the pump pulse has already become negligible, and rewrite
\begin{multline}
\sigma_{\text{out},\text{in}}\left(\omega,t_{\mathrm{pr}}\right)=\int_{-\infty}^{t_{\mathrm{fin}}}\mathrm{e}^{\mathrm{i}\left(\omega+\mathrm{i}\lambda\right)\left(t-t_{\mathrm{pr}}\right)}\sigma_{\text{out},\text{in}}\left(t,t_{\mathrm{pr}}\right)dt\\
+\sigma_{\text{out},\text{in}}^{\text{a.p.}}\left(\omega,t_{\mathrm{fin}},t_{\mathrm{pr}}\right),\label{eq:sig_w_integ_ap}
\end{multline}
where the second term, $\sigma_{\text{out},\text{in}}^{\text{a.p.}}$,
accounts for the contribution accumulated after the pump pulse has
become negligible. This term naturally splits as
\begin{multline}
\sigma_{\text{out},\text{in}}^{\text{a.p.}}\left(\omega,t_{\mathrm{fin}},t_{\mathrm{pr}}\right)=\sigma_{\text{out},\text{in}}^{\left(1\right),\text{a.p.}}\left(\omega,t_{\mathrm{fin}},t_{\mathrm{pr}}\right)+\\
+\sigma_{\text{out},\text{in}}^{\left(2\right),\text{a.p.}}\left(\omega,t_{\mathrm{fin}},t_{\mathrm{pr}}\right),
\end{multline}
and can be evaluated exploiting the fact that the Hamiltonian becomes
stationary, once the pump pulse has become negligible, and reverts
to its equilibrium form. One then obtains \citep{eskandari2025dynamical}
\begin{multline}
\sigma_{\text{out},\text{in}}^{\left(1\right),\text{a.p.}}\left(\omega,t_{\mathrm{fin}},t_{\mathrm{pr}}\right)=-\frac{\mathrm{i}e}{\hbar\mathcal{V}}\mathrm{e}^{\mathrm{i}\left(\omega+\mathrm{i}\lambda\right)\left(t_{\mathrm{fin}}-t_{\mathrm{pr}}\right)}\times\\
\times\sum_{\boldsymbol{k}}\left\{ W_{\boldsymbol{k}}\left(\omega,t_{\mathrm{fin}}\right)+\mathrm{Tr}\left[Q_{\boldsymbol{k}}\left(\omega,t_{\mathrm{fin}}\right)\cdot S_{\boldsymbol{k}}\left(t_{\mathrm{fin}},t_{\mathrm{pr}}\right)\right]\right\} ,\label{eq:sig_ap_1}
\end{multline}
where the auxiliary matrices read 
\begin{multline}
Q_{\boldsymbol{k}}\left(\omega,t_{\mathrm{fin}}\right)=\\
=\mathrm{i}\left\{ P_{\boldsymbol{k}}^{\dagger}\left(t_{\mathrm{fin}}\right)\cdot\left(J_{\boldsymbol{k},\text{out}}\left(t_{\mathrm{fin}}\right)\circ\bar{w}_{\boldsymbol{k}}\left(\omega\right)\right)\cdot P_{\boldsymbol{k}}\left(t_{\mathrm{fin}}\right)\right\} \circ\Delta F_{\boldsymbol{k}},\label{eq:Qwt}
\end{multline}
\begin{equation}
S_{\boldsymbol{k}}\left(t_{\mathrm{fin}},t_{\mathrm{pr}}\right)=-Y_{\boldsymbol{k}}\left(t_{\mathrm{fin}}\right)+Y_{\boldsymbol{k}}\left(t_{\mathrm{pr}}\right)-X_{\boldsymbol{k}}\left(t_{\mathrm{pr}}\right),\label{eq:Swt}
\end{equation}
\begin{multline}
W_{\boldsymbol{k}}\left(\omega,t_{\mathrm{fin}}\right)=-\frac{e}{\hbar}\left\{ \text{Tr}\left(J_{\boldsymbol{k},\text{out}}\left(t_{\mathrm{fin}}\right)\circ\bar{w}_{\boldsymbol{k}}\left(\omega\right)\right)\cdot\right.\\
\left.\left[\eta_{\boldsymbol{k},\text{in}},\rho_{\boldsymbol{k}}\left(t_{\mathrm{fin}}\right)\circ\bar{w}_{\boldsymbol{k}}^{\text{T}}\left(\omega\right)\right]\right\} ,
\end{multline}
with $\bar{w}_{\boldsymbol{k},n_{1},n_{2}}\left(\omega\right)=\frac{1}{\omega+\omega_{\boldsymbol{k},n_{1},n_{2}}+\mathrm{i}\lambda}$,
$\omega_{\boldsymbol{k},n_{1},n_{2}}=\left(\varepsilon_{\boldsymbol{k},n_{1}}-\varepsilon_{\boldsymbol{k},n_{2}}\right)/\hbar$,
and the superscript $\text{T}$ denoting transposition. Similarly,
one finds 
\begin{multline}
\sigma_{\text{out},\text{in}}^{\left(2\right),\text{a.p.}}\left(\omega,t_{\mathrm{fin}},t_{\mathrm{pr}}\right)=\frac{e\mathrm{i}}{\mathcal{V}}\mathrm{e}^{\mathrm{i}\left(\omega+\mathrm{i}\lambda\right)\left(t_{\mathrm{fin}}-t_{\mathrm{pr}}\right)}\times\\
\times\sum_{\boldsymbol{k}}\text{Tr}\left\{ \left(\left(\frac{\delta J_{\boldsymbol{k}}}{\delta A}\right)_{\text{out},\text{in}}\left(t_{\mathrm{fin}}\right)\circ\bar{w}_{\boldsymbol{k}}\left(\omega\right)\right).\rho_{\boldsymbol{k}}\left(t_{\mathrm{fin}}\right)\right\} \cdot\label{eq:sig_ap_2}
\end{multline}

For the evaluation of these expressions, it is important to notice
that both the hopping and dipole matrices, together with all their
derivatives with respect to $\boldsymbol{k}$ (and hence also $J_{\boldsymbol{k},\text{out}},\,\eta_{\boldsymbol{k},\text{in}}$,
and $\left(\frac{\delta J_{\boldsymbol{k}}}{\delta A}\right)_{\text{out},\text{in}}$),
return to their equilibrium values at $t_{\mathrm{fin}}$.

\subsection{After-pump optical conductivity and dissipative effects}\label{subsec:Optical-conductivity-SPDM}

Once the pump pulse has become negligible, the Hamiltonian relaxes
to its equilibrium form and becomes stationary. This fact enables
a considerable simplification: the optical conductivity, which in
general is a two-time function, can now be expressed solely in terms
of the SPDM, which depends only on one time. Such a simplification
is clearly not available if the probe pulse is applied while the pump
pulse is still active, i.e., when $t_{\mathrm{pr}}$ lies within the
pumping time interval. The stationary-Hamiltonian regime has already
been addressed in the previous section: in Eq.~\ref{eq:sig_w_integ_ap},
we reported the contribution accumulated after the pump pulse has
become negligible.

To derive the post-pump optical conductivity, given that $t_{\mathrm{pr}}$
lies outside the pumping time interval, having the full SPDM, the
Fourier transformation of the optical conductivity can be performed
analytically and one can set $t_{\mathrm{fin}}=t_{\mathrm{pr}}$ in
Eq.~\ref{eq:sig_w_integ_ap}, which reduces to zero the first term
on its right-hand side. Having set $t_{\mathrm{fin}}=t_{\mathrm{pr}}$
in Eqs.~\eqref{eq:sig_ap_1} and~\eqref{eq:sig_ap_2}, and using
the identity $P_{\boldsymbol{k}}^{\dagger}\left(t\right)\cdot P_{\boldsymbol{k}}\left(t\right)=\boldsymbol{1}$,
we arrive, after straightforward algebra, at (see App. \ref{sec:Derivation-sigma_rho})
\begin{equation}
\sigma_{\text{out},\text{in}}\left(\omega,t_{\mathrm{pr}}\right)=\sigma_{\text{out},\text{in}}^{\left(1\right)}\left(\omega,t_{\mathrm{pr}}\right)+\sigma_{\text{out},\text{in}}^{\left(2\right)}\left(\omega,t_{\mathrm{pr}}\right),\label{eq:sigma-rho}
\end{equation}
with 
\begin{multline}
\sigma_{\text{out},\text{in}}^{\left(1\right)}\left(\omega,t_{\mathrm{pr}}\right)=-\frac{\mathrm{i}e}{\hbar\mathcal{V}}\sum_{\boldsymbol{k}}\left\{ W_{\boldsymbol{k}}\left(\omega,t_{\mathrm{pr}}\right)+\right.\\
\left.+\mathrm{i}e\mathrm{Tr}\left(\left[D_{\boldsymbol{k},\text{in}}\left(t_{\mathrm{pr}}\right),J_{\boldsymbol{k},\text{out}}\left(t_{\mathrm{pr}}\right)\circ\bar{w}_{\boldsymbol{k}}\left(\omega\right)\right]\cdot\rho_{\boldsymbol{k}}\left(t_{\mathrm{pr}}\right)\right)\right\} ,\label{eq:sigma-rho-1}
\end{multline}
\begin{align}
 & \sigma_{\text{out},\text{in}}^{\left(2\right)}\left(\omega,t_{\mathrm{pr}}\right)=\frac{\mathrm{i}e}{\mathcal{V}}\times\nonumber \\
 & \times\sum_{\boldsymbol{k}}\text{Tr}\left\{ \left(\left(\frac{\delta J_{\boldsymbol{k}}}{\delta A}\right)_{\text{out},\text{in}}\left(t_{\mathrm{pr}}\right)\circ\bar{w}_{\boldsymbol{k}}\left(\omega\right)\right)\cdot\rho_{\boldsymbol{k}}\left(t_{\mathrm{pr}}\right)\right\} .\label{eq:sigma-rho-2}
\end{align}

An essential advantage of this formalism is that it allows one to
consistently incorporate dissipative effects. In particular, one can
add a phenomenological Markovian damping directly to the SPDM dynamics,
thereby investigating its impact on the optical response. The essential
requirement is that the damping timescale remains long compared to
the oscillation period of the probe pulse ($\sim2\pi/\omega$), so
that the Hamiltonian remains effectively stationary. Considering that
the decoherence and/or relaxation time scales are of the orders of
several tens to hundreds of femtoseconds in many experimental setups
(see Ref.~\citep{inzani2023field} and the references there in),
our phenomenological approach is valid in a wide range of cases.

The modified equation of motion for the SPDM then reads 
\begin{equation}
\partial_{t}\rho_{\boldsymbol{k}}\left(t\right)=-\frac{\mathrm{i}}{\hbar}\left[\Xi_{\mathbf{k}}\left(t\right),\rho_{\boldsymbol{k}}\left(t\right)\right]-\Upsilon\circ\left(\rho_{\boldsymbol{k}}\left(t\right)-\rho_{\boldsymbol{k}}\left(t_{\mathrm{ini}}\right)\right),\label{eq:EOM_rho}
\end{equation}
where $\rho_{\boldsymbol{k}}\left(t_{\mathrm{ini}}\right)=F_{\boldsymbol{k}}$
is the initial equilibrium distribution, and the matrix $\Upsilon$
encodes the phenomenological damping coefficients governing the relaxation
processes.

In addition to enabling the straightforward incorporation of phenomenological
damping, using SPDM in regimes where this is valid reduces computational
cost relative to full DPOA calculations by a noticeable factor, which
can reach an order of magnitude, depending on the final probe time.
It is worth noting that the full DPOA calculations, in turn, are much
more efficient (by one or even two orders of magnitude) than some
other standard approaches, such as TD-DFT calculations \citep{inzani2023field,eskandari2024time,eskandari2024generalized}.

\subsection{Magneto-optical Kerr rotation}

The magneto-optical Kerr effect provides a powerful probe of symmetry
breaking in solids, and is usually expressed in terms of the optical
conductivity tensor, $\boldsymbol{\sigma}\left(\omega,t_{\mathrm{pr}}\right)$,
or equivalently, the dielectric tensor, $\boldsymbol{\epsilon}\left(\omega,t_{\mathrm{pr}}\right)$.
The latter is computed in terms of the former as \citep{eskandari2024generalized}
\begin{equation}
\boldsymbol{\epsilon}\left(\omega,t_{\mathrm{pr}}\right)=\boldsymbol{1}+\tfrac{\mathrm{i}}{\omega\epsilon_{0}}\boldsymbol{\sigma}\left(\omega,t_{\mathrm{pr}}\right),\label{eq:eps_sigma}
\end{equation}
where $\epsilon_{0}$ denotes the vacuum permittivity. In the simplest
theoretical treatments, one evaluates the Kerr rotation angle from
the longitudinal component, $\sigma_{xx}$, together with the transverse
component, $\sigma_{xy}$, under the assumption that the latter is
simply antisymmetric, i.e., $\sigma_{xy}=-\sigma_{yx}$, after the
breakdown of time-reversal symmetry (TRS). Here, the $x$-axis is
chosen to be along the incoming probe pulse polarization, and the
$y$-axis is chosen in such a way that in a right-handed system the
propagation direction of the incoming probe pulse is along $-z$.
However, in more general situations---such as in crystals with  linear
birefringence---the transverse components of the optical conductivity
may remain finite even in the absence of TRS breaking. In that case,
they are symmetric rather than antisymmetric. To address this, we
explicitly isolate the odd (antisymmetric) component of the transverse
optical conductivity, 
\begin{equation}
\sigma_{xy}^{\text{odd}}=\tfrac{1}{2}\left(\sigma_{xy}-\sigma_{yx}\right),
\end{equation}
and employ this quantity to evaluate the Kerr response. For a probe
pulse incident perpendicularly to the sample surface, the  magneto-optical
(polar) Kerr rotation angle, $\theta_{K}\left(\omega,t_{\mathrm{pr}}\right)$,
is then obtained as (see App.~\ref{sec:Computing-Polar-Kerr}, and
substitute Eq.~\ref{eq:eps_sigma} in Eq.~\ref{eq:Kerr_app_eps})
\begin{multline}
\theta_{K}\left(\omega,t_{\mathrm{pr}}\right)=\\
=\Re\frac{\sigma_{xy}^{\text{odd}}\left(\omega,t_{\mathrm{pr}}\right)}{\sigma_{d}\left(\omega,t_{\mathrm{pr}}\right)\sqrt{1+\tfrac{\mathrm{i}}{\omega\epsilon_{0}}\sigma_{d}\left(\omega,t_{\mathrm{pr}}\right)}},\label{eq:Kerr}
\end{multline}
where $\sigma_{d}=\frac{1}{2}\left(\sigma_{xx}+\sigma_{yy}\right)$.

\section{Two-band model}\label{sec:Numerical-studies}

In this section, we apply the formalism developed above to a minimal
two-band model designed to capture the essential physics of pump--probe
Kerr rotation angle dynamics in systems with spin--orbit coupling
(SOC) and broken TRS. This simplified model, which includes nearest-neighbor
hopping, Rashba SOC, and Zeeman splitting, while not intended to reproduce
the detailed electronic structure of a specific material, enables
us to disentangle the microscopic mechanisms underlying the transient
optical conductivity and the ensuing Kerr rotation. The results presented
here illustrate the distinct features that emerge in and out of equilibrium
and establish the connection between the microscopic electronic processes
and experimentally measurable observables.

\begin{figure*}
\centering{}\includegraphics{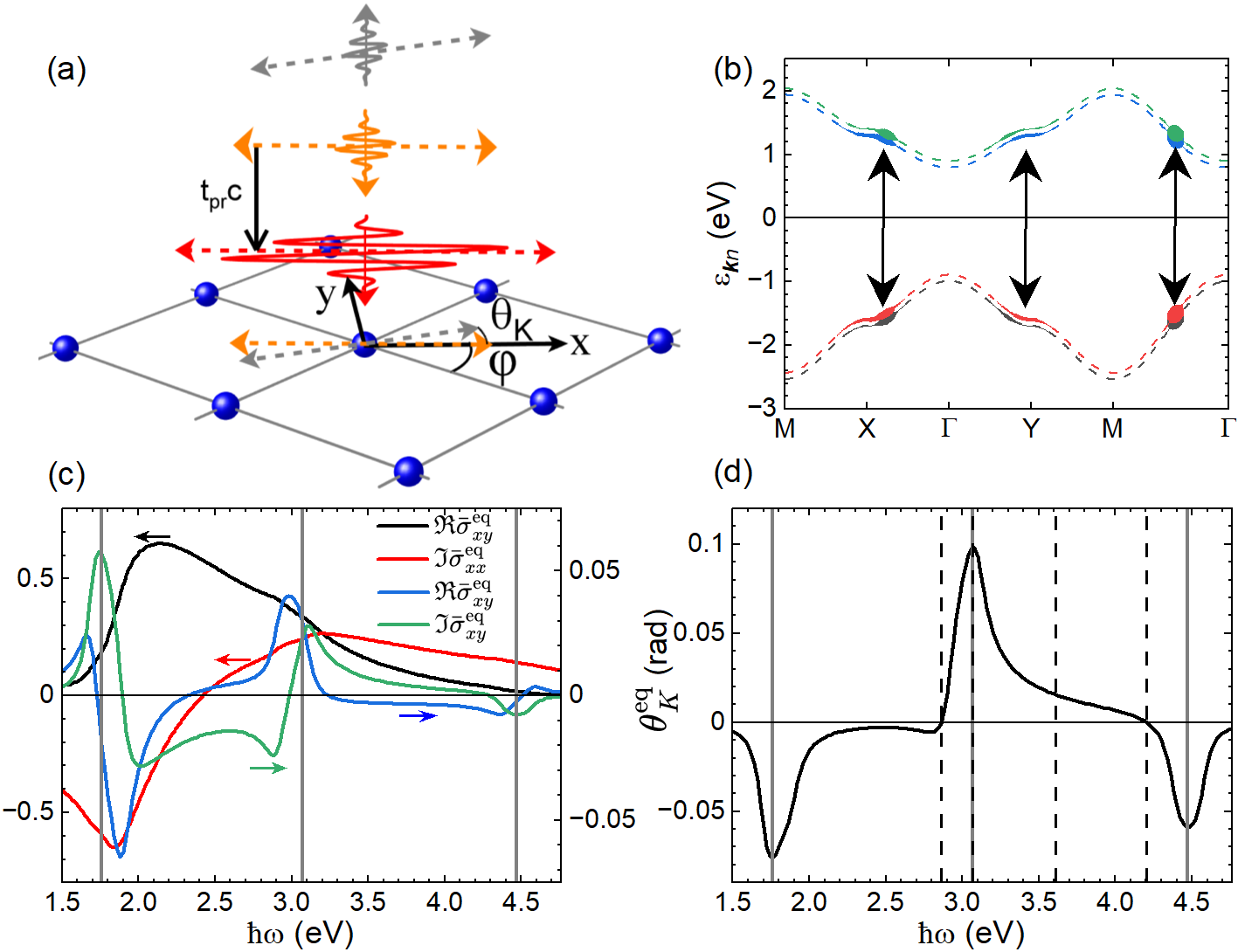} \caption{(a) Schematic representation of the system, the pump pulse, and the
time-delayed probe pulse together with its reflected component, showing
the Kerr rotation of the reflected probe pulse polarization. The probe
pulse arrives with a delay $t_{\mathrm{pr}}$ with respect to the
pump pulse. The angle between the probe pulse polarization and the
lattice vector is $\varphi$, which is set to $\varphi=0$ in our
calculations. (b) Band structure of the equilibrium Hamiltonian including
Rashba SOC and Zeeman splitting. The thickness of the solid lines
on top of the dashed lines indicates the post-pump excitations, hole
or electron, in VB or CB, respectively, for the pump pulse photon
energy of $\unit[2.86]{eV}$. The double arrows mark the resonant
inter-band energy gaps. (c) Real and imaginary parts of the equilibrium
optical conductivities $\bar{\sigma}_{xx}^{\text{eq}}\left(\omega\right)$
and $\bar{\sigma}_{xy}^{\text{eq}}\left(\omega\right)$. Vertical
gray solid lines mark the local extrema of $\theta_{K}^{\text{eq}}\left(\omega\right)$,
see panel (d). (d) Equilibrium Kerr rotation angle, $\theta_{K}^{\text{eq}}\left(\omega\right)$.
Vertical gray lines correspond to the local extrema of $\theta_{K}^{\text{eq}}\left(\omega\right)$,
and vertical dashed lines indicate the pump photon energies used in
the time-dependent simulations.}\label{fig:system_eq}
\end{figure*}

\begin{figure*}
\centering{}\includegraphics{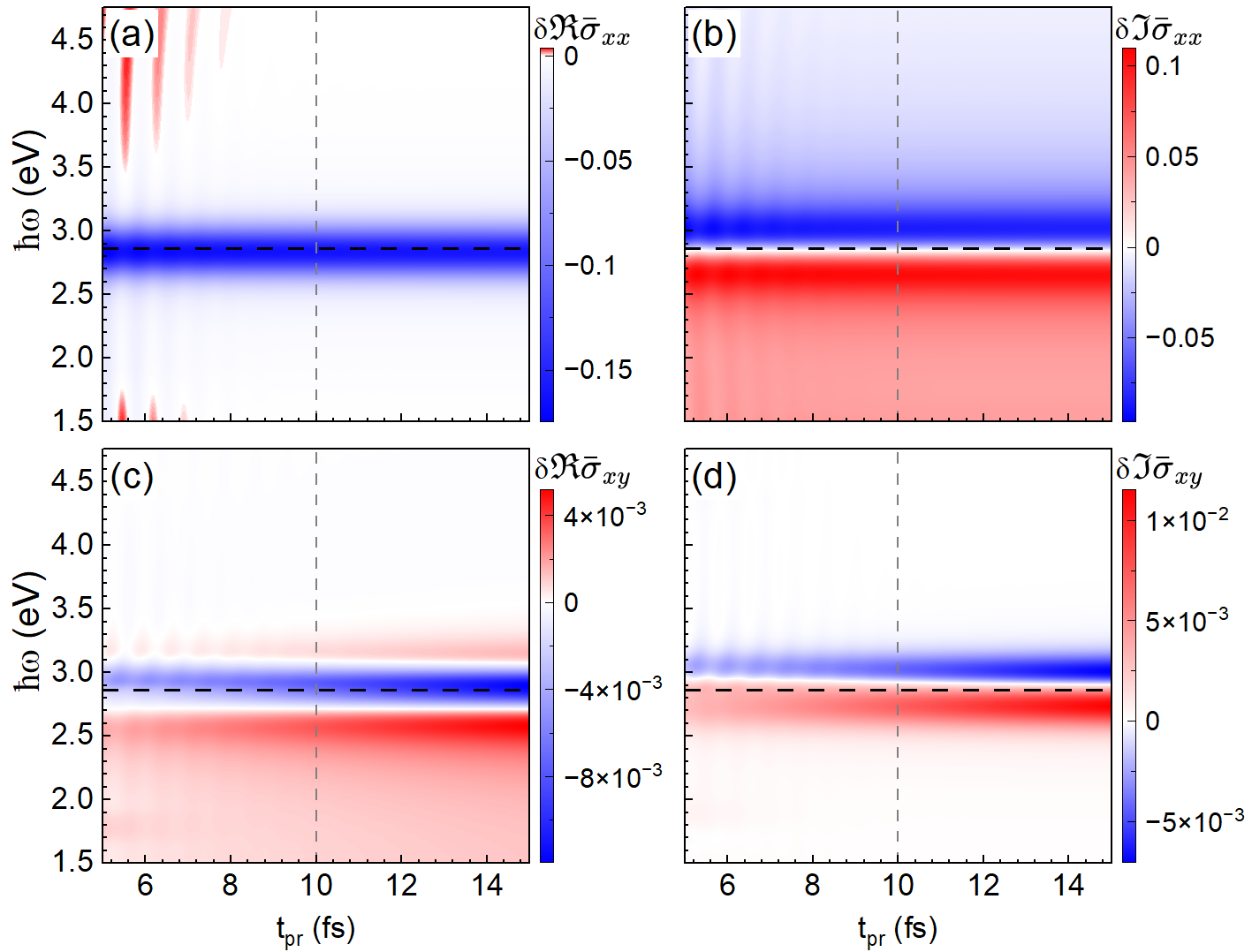} \caption{(a--d) Real and imaginary parts of $\delta\bar{\sigma}_{xx}(\omega,t_{\mathrm{pr}})$
and $\delta\bar{\sigma}_{xy}(\omega,t_{\mathrm{pr}})$ as functions
of $\omega$ and $t_{\mathrm{pr}}$. The horizontal dashed lines mark
the pump pulse frequency $\hbar\omega_{\text{pu}}=\unit[2.86]{eV}$,
while the vertical dashed lines indicate the probe pulse delay equal
to the pump pulse FWHM, $t_{\mathrm{pr}}=\tau_{\mathrm{pu}}=\unit[10]{fs}$.}\label{fig:d_sigma}
\end{figure*}

\begin{figure*}
\centering{}\includegraphics{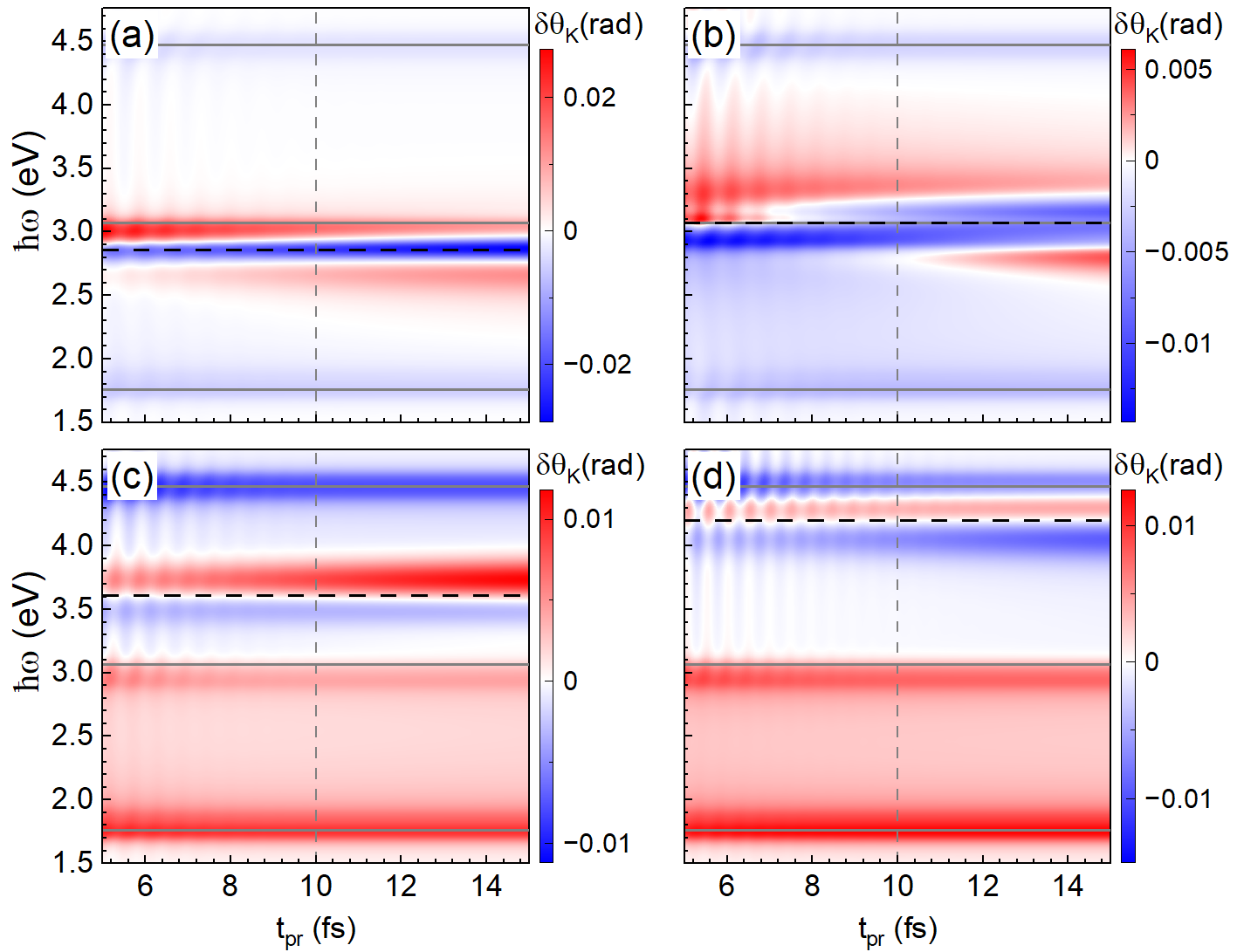} \caption{(a--d) The evolution of the Kerr rotation angle, $\delta\theta_{K}(\omega,t_{\mathrm{pr}})$,
for four different pump pulse frequencies indicated in Fig.~\ref{fig:system_eq}(d).
Horizontal dashed lines correspond to the respective pump pulse frequencies.
Horizontal gray lines correspond to the local extrema of $\theta_{K}^{\text{eq}}\left(\omega\right)$,
see Fig.~\ref{fig:system_eq} (d).}\label{fig:d_Kerr}
\end{figure*}

\subsection{Model, computational details and equilibrium optical response}

The system consists of a two-orbital tight-binding model on a square
lattice with lattice constant set to $a=\unit[1]{nm}$ for convenience.
Each orbital accommodates two spin states, resulting in a total of
four bands. The onsite energies and nearest-neighbor hoppings within
a single spin subspace are defined in the following.

We take the valence band (VB) and conduction band (CB) to originate
from two localized Wannier states with onsite energies $\tilde{T}_{\boldsymbol{R}=0,1,1}^{\text{hop}}=-\unit[1.65]{eV}$
and $\tilde{T}_{\boldsymbol{R}=0,2,2}^{\text{hop}}=\unit[1.35]{eV}$,
respectively. The diagonal nearest-neighbor hoppings are $\tilde{T}_{\mathbf{R}=\mathbf{a},1,1}^{\text{hop}}=\unit[0.2]{eV}$
and $\tilde{T}_{\mathbf{R}=\mathbf{a},2,2}^{\text{hop}}=\unit[-0.15]{eV}$,
while the off-diagonal hoppings are $\tilde{T}_{\mathbf{R}=\mathbf{a},1,2}^{\text{hop}}=\tilde{T}_{\mathbf{R}=\mathbf{a},2,1}^{\text{hop}}=\unit[-0.1]{eV}$.
Here, $\tilde{T}_{\boldsymbol{R},\nu,\nu'}^{\text{hop}}$ denotes
the hopping matrix between two Wannier states $\nu$ and $\nu'$ centered
at sites separated by $\boldsymbol{R}$, and $\mathbf{a}\in\{a(\pm1,0),a(0,\pm1)\}$.
For bulk calculations, the BZ is sampled as a $128\times128$ $\boldsymbol{k}$-grid
centered at $\Gamma$. The damping factor used in evaluating the optical
conductivity is set to $\lambda=\unit[0.1]{PHz}$. Fourier transforming
$\tilde{T}_{\boldsymbol{R}}^{\text{hop}}$ yields the momentum-space
hopping matrix $\tilde{T}_{\boldsymbol{k}}^{\text{hop}}$~\citep{eskandari2024time,eskandari2024generalized}.

This minimal model allows us to explore generic phenomena that can
arise in realistic pump--probe experiments, while avoiding complications
related to material-specific details. The underlying formalism, however,
remains general and can be applied to any lattice structure or multi-band
systems.

A key ingredient for the emergence of Kerr rotation is the presence
of SOC. We include a Rashba SOC term of the form 
\begin{equation}
\tilde{T}_{\boldsymbol{k}}^{\text{Rashba}}=\alpha_{\text{Rashba}}\left[\sin(k_{x}a)\varsigma_{y}^{\text{spin}}-\sin(k_{y}a)\varsigma_{x}^{\text{spin}}\right]\otimes\varsigma_{y}^{\text{orb}},
\end{equation}
where $\varsigma_{x/y}^{\text{spin/orb}}$ are the Pauli matrices
acting on the spin and orbital subspaces, respectively, and we set
$\alpha_{\text{Rashba}}=\unit[0.1]{eV}$. The Rashba term alone preserves
TRS. To break TRS, we include a Zeeman term 
\begin{equation}
\tilde{T}_{\boldsymbol{k}}^{\text{Zeeman}}=\alpha_{\text{Zeeman}}\varsigma_{z}^{\text{spin}},
\end{equation}
with $\alpha_{\text{Zeeman}}=\unit[0.05]{eV}$. The total equilibrium
Hamiltonian in the localized Wannier basis is therefore given by 
\begin{equation}
\tilde{T}_{\boldsymbol{k}}=\tilde{T}_{\boldsymbol{k}}^{\text{hop}}+\tilde{T}_{\boldsymbol{k}}^{\text{Rashba}}+\tilde{T}_{\boldsymbol{k}}^{\text{Zeeman}}.
\end{equation}

We set the chemical potential and the temperature to zero. As such,
we have two VBs and two CBs.

We simulate a pump--probe experiment where the system is driven by
a linearly polarized pump pulse described by a vector potential $\boldsymbol{A}_{\mathrm{pu}}(t)=A_{\mathrm{pu}}(t)\hat{\boldsymbol{u}}_{\text{pu}}$,
with 
\begin{equation}
A_{\mathrm{pu}}(t)=A_{0}e^{-(4\ln2)t^{2}/\tau_{\mathrm{pu}}^{2}}\sin(\omega_{\mathrm{pu}}t),
\end{equation}
where $\tau_{\mathrm{pu}}$ is the full width at half maximum (FWHM)
of the pulse envelope. We set $\tau_{\mathrm{pu}}=\unit[10]{fs}$
and the pump pulse amplitude $A_{0}=\unitfrac[0.5]{V}{nm\,fs}$. A
time-delayed probe pulse is applied with polarization $\hat{\boldsymbol{u}}_{\text{in}}=\hat{\boldsymbol{u}}_{\text{pu}}=\hat{x}$.
The polarization of the reflected probe pulse is changed by the Kerr
rotation angle, $\theta_{K}(\omega,t_{\mathrm{pr}})$.

Figure~\ref{fig:system_eq}(a) schematically shows the overall geometry
of the system, the incident pump and probe pulses, and the Kerr rotation
of the reflected probe pulse.

Post-pump excitations occur mainly at $\boldsymbol{k}$-points where
the inter-band gap matches the pump photon energy $\hbar\omega_{\text{pu}}=\unit[2.86]{eV}$,
as shown in Fig.~\ref{fig:system_eq}(b), on top of the system bands
along a specific path in the BZ. Along the $\Gamma$--Y path, small
excitations also appear due to SOC, which couples spin and orbital
degrees of freedom. Without SOC, no excitations would occur there
for the chosen pump polarization, because $\eta_{\boldsymbol{k},x}=0$
for the $\boldsymbol{k}$-points along $\Gamma$--Y for this model~\citep{eskandari2024time}.

\begin{figure*}
\centering{}\includegraphics{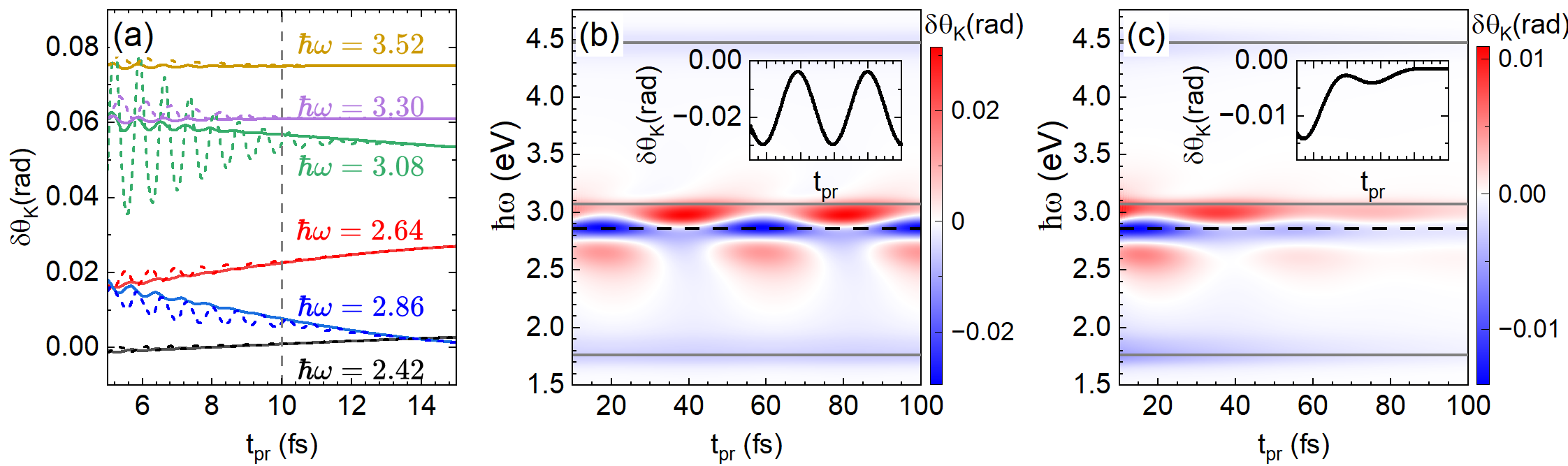} \caption{(a) Benchmark of the exact results for $\delta\theta_{K}(\omega,t_{\mathrm{pr}})$
against those obtained from the SPDM approach without damping ($\Upsilon=0$
in Eq.~\ref{eq:EOM_rho}) for several probe pulse frequency cuts.
The probe photon energy values of each cut, $\hbar\omega$, are given
in eV. An offset of $\unit[0.015]{rad}$ has been applied on increasing
the probe photon energies. The pump photon energy is $\hbar\omega_{\text{pu}}=\unit[2.86]{eV}$.
(b) $\delta\theta_{K}(\omega,t_{\mathrm{pr}})$ at long delays without
damping; the inset shows a cut at the resonant frequency $\omega=\omega_{\text{pu}}$.
(c) Same as (b) but including damping in SPDM with $\Upsilon_{nm}=\delta_{nm}\tfrac{\lambda}{4}+(1-\delta_{nm})\tfrac{\lambda}{2}$.
The horizontal solid and dashed lines in panels (b) and (c) are the
same as those in Fig.~\ref{fig:d_Kerr} (a).}\label{fig:d_Kerr-long-t}
\end{figure*}

\begin{figure*}
\centering{}\includegraphics{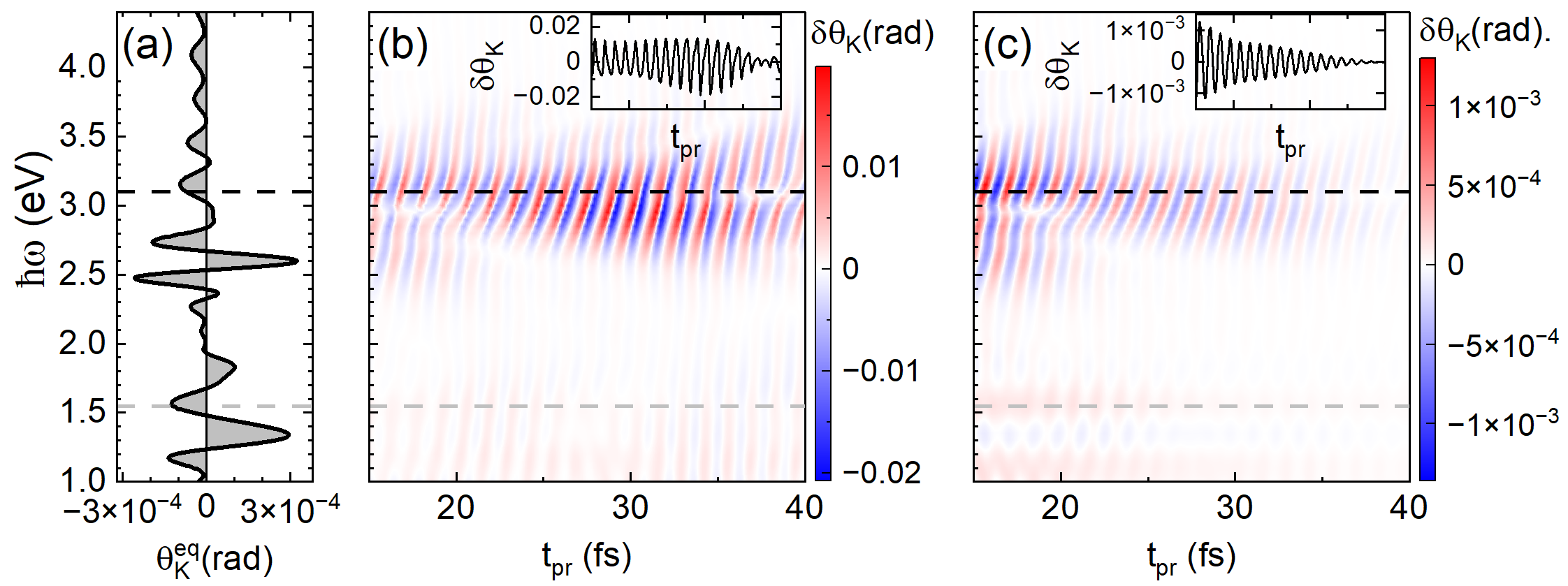} \caption{(a) Equilibrium Kerr rotation $\theta_{K}^{\text{eq}}\left(\omega\right)$
of weakly spin-polarized germanium. (b) $\delta\theta_{K}(\omega,t_{\mathrm{pr}})$
at long delays without damping; the inset shows a cut at the two-photon
resonant frequency $\omega=2\omega_{\text{pu}}$. (c) Same as (b)
but including damping in SPDM with $\Upsilon_{nm}=\delta_{nm}\tfrac{\lambda}{4}+(1-\delta_{nm})\tfrac{\lambda}{2}$.
The horizontal gray and black dashed lines in all panels mark the
one and two-photon resonances, $\hbar\omega=\hbar\omega_{\text{pu}}=\unit[1.55]{eV}$
and $\hbar\omega=2\hbar\omega_{\text{pu}}=\unit[3.10]{eV}$, respectively.}\label{fig:germanium}
\end{figure*}

When the probe pulse arrives well before the pump pulse (equivalently,
when no pump pulse is applied), the optical response corresponds to
the equilibrium case, and is denoted by the superscript ``eq'' (the
expression for the analytical equilibrium optical conductivity is
reported in Refs.~\citep{eskandari2024generalized,eskandari2025dynamical}).
For plotting convenience, all optical conductivities are reported
in the figures in a dimensionless form $\bar{\sigma}=\sigma/(\omega\epsilon_{0})$.
The equilibrium components $\bar{\sigma}_{xx}^{\text{eq}}(\omega)$
and $\bar{\sigma}_{xy}^{\text{eq}}(\omega)$ are shown in Fig.~\ref{fig:system_eq}(c).
The real part, $\Re\bar{\sigma}_{xx}^{\text{eq}}$, is related to
absorption and decreases outside the band-gap regions. The real and
imaginary parts of $\bar{\sigma}$ are related through the Kramers--Kronig
(KK) relations, implying that local extrema in one correspond to dispersive
features in the other. We dub such features as KK-counterparts.

The equilibrium Kerr rotation angle, $\theta_{K}^{\text{eq}}(\omega)$,
shown in Fig.~\ref{fig:system_eq}(d), exhibits local extrema around
the same frequencies where the KK-counterpart features appear in $\Re\bar{\sigma}_{xy}^{\text{eq}}(\omega)$
and $\Im\bar{\sigma}_{xy}^{\text{eq}}(\omega)$. However, the dominant
feature cannot be inferred a priori due to the nonlinear dependence
of the Kerr rotation on the optical conductivity tensor, see Eq.~\ref{eq:Kerr}.
The vertical dashed lines indicate the four pump pulse frequencies
considered in the following simulations, one of which coincides with
a local maximum in $\theta_{K}^{\text{eq}}(\omega)$. Unless otherwise
specified, the pump pulse frequency for the toy model is $\hbar\omega_{\text{pu}}=\unit[2.86]{eV}$,
the frequency at which we have $\theta_{K}^{\text{eq}}(\omega_{\text{pu}})=0$.

\subsection{Transient optical conductivity and Kerr rotation}

Figure~\ref{fig:d_sigma} shows the temporal evolution of the changes
in optical conductivity, $\delta\bar{\sigma}_{\alpha\beta}(\omega,t_{\mathrm{pr}})=\bar{\sigma}_{\alpha\beta}(\omega,t_{\mathrm{pr}})-\bar{\sigma}_{\alpha\beta}^{\text{eq}}(\omega)$,
which is dubbed differential optical conductivity. During the pump
pulse application ($t_{\mathrm{pr}}\lesssim\tau_{\mathrm{pu}}$),
both real and imaginary parts exhibit rapid oscillations with frequency
$\sim2\omega_{\text{pu}}$. After pumping the system ($t_{\mathrm{pr}}\gg\tau_{\mathrm{pu}}$),
these fast oscillations disappear and the real parts of the differential
optical conductivities remain negative near the resonant frequency
$\omega\approx\omega_{\text{pu}}$ (the blue stripes in Figs.~\ref{fig:d_sigma}
(a) and (c)). This behavior reflects a reduction in absorption caused
by photo-excited carriers that block further probe-induced inter-band
transitions, the so-called Pauli blocking~\citep{eskandari2024generalized},
which is also dubbed state blocking \citep{Koopmans:2000aa,Oppeneer:2004aa}.
The imaginary parts of the optical conductivities display KK-counterpart
features relative to their real parts.

The evolution of the Kerr rotation, $\delta\theta_{K}(\omega,t_{\mathrm{pr}})=\theta_{K}(\omega,t_{\mathrm{pr}})-\theta_{K}^{\text{eq}}(\omega)$,
for four distinct pump pulse frequencies is presented in Fig.~\ref{fig:d_Kerr}.
During the pump pulse time interval ($t_{\mathrm{pr}}\lesssim\tau_{\mathrm{pu}}$),
$\delta\theta_{K}$ exhibits fast oscillations. These fast oscillations
vanish when the pump pulse becomes negligible ($t_{\mathrm{pr}}\gg\tau_{\mathrm{pu}}$).
Near-resonant features appear around $\omega\approx\omega_{\text{pu}}$,
resembling those in Fig.~\ref{fig:d_sigma}. The detailed form of
these near-resonant features---whether stripe-like {[}as in Figs.~\ref{fig:d_Kerr}
(a) and (b){]} or its KK counterpart {[}as in Figs.~\ref{fig:d_Kerr}
(c) and (d){]}---depends on the complex interplay of real and imaginary
parts of $\sigma_{xx}$ and $\sigma_{xy}$ in Eq.~\ref{eq:Kerr},
and therefore, is different for different pump pulse frequencies.

Interestingly, additional features emerge at the frequencies where
the equilibrium Kerr rotation exhibits local extrema {[}vertical gray
lines in Fig.~\ref{fig:system_eq}(d){]}. This can be understood
by considering a simplified ratio $f(\omega)=g(\omega)/h(\omega)$
that mimics the Kerr functional dependence. Upon pumping, $g\to g^{\text{eq}}+\delta g$
and $h\to h^{\text{eq}}+\delta h$, which results in a variation,
$f\to f^{\text{eq}}+\delta f$ , which up to the first order reads
$\delta f=\delta g/h^{\text{eq}}-(\delta h/h^{\text{eq}})f^{\text{eq}}$.
Thus, even when $\delta g$ and $\delta h$ are small off resonance,
the product with $f^{\text{eq}}$ produces additional structures at
its local extrema.

\subsection{Long-time dynamics and damping effects}

In many experiments, the long-delay regime ($t_{\mathrm{pr}}\gg\tau_{\mathrm{pu}}$)
of the Kerr response is of particular interest. Calculations in such
regime become computationally more feasible within the SPDM framework,
which, as discussed in Sec.~\ref{subsec:Optical-conductivity-SPDM},
is valid only after the pump pulse has become negligible. Fig.~\ref{fig:d_Kerr-long-t}(a)
benchmarks the exact $\delta\theta_{K}$ results against SPDM simulations
without damping ($\Upsilon=0$ in Eq.~\ref{eq:EOM_rho}). The agreement
improves systematically as $t_{\mathrm{pr}}$ exceeds the pump pulse
duration, confirming the applicability of the SPDM approach.

Figure~\ref{fig:d_Kerr-long-t}(b) displays $\delta\theta_{K}(\omega,t_{\mathrm{pr}})$
for long delays without damping. A slow oscillation in time appears,
whose frequency corresponds to $\tfrac{2\pi\hbar}{2\alpha_{\text{Zeeman}}}$.
This beating pattern originates from the interference between off-diagonal
SPDM elements $\rho_{\boldsymbol{k}nm}(t)$ (the \emph{coherences})
oscillating with slightly different inter-band energy gaps, which
in our case comes from the Zeeman splitting. In real materials, similar
beating effects are expected due to small band splittings caused by
SOC, spin-polarization, Zeeman terms, etc.

Finally, Fig.~\ref{fig:d_Kerr-long-t}(c) shows the impact of damping,
implemented phenomenologically in SPDM via $\Upsilon_{nm}=\delta_{nm}\tfrac{\lambda}{4}+(1-\delta_{nm})\tfrac{\lambda}{2}$,
where off-diagonal elements decay faster than the diagonal ones \citep{mukamel1995principles}.
Damping reduces the overall amplitude of the Kerr rotation angle,
and gradually suppresses oscillations as $t_{\mathrm{pr}}$ increases.

\section{Weakly spin-polarized germanium}\label{sec:germanium}

In this section, we present our results for weakly spin-polarized
germanium, illustrating the capability of the framework to compute
the Kerr rotation in a real material while accounting for all the
difficulties and complexities associated with a realistic band structure.

\subsection{Weakly spin-polarized germanium and pump-probe setup}

We consider weakly spin-polarized germanium (magnetic moment per cell
$\approx\unit[6\times10^{-5}]{\mu_{B}}$) to break TRS and enable
the measurement of a finite Kerr rotation in a material that intrinsically
possesses finite SOC. The spin polarization can be in principle achieved
experimentally through standard methods, such as exploiting proximity
effects \citep{PhysRevB.70.033201}. The band structure was obtained
with the Elk code \citep{elk_code} in the presence of an \emph{ad
hoc} magnetic field, and the hopping parameters ($\tilde{T}_{\mathbf{k}}$)
and dipole matrix elements ($\tilde{\mathbf{D}}_{\mathbf{k}}$) were
computed using the Wannier90 code \citep{pizzi2020wannier90}, as
in Ref.~\citep{inzani2023field}. We focused on the closed manifold
of 16 $sp^{3}$ bands surrounding the chemical potential, which lies
within the main band gap of $\approx\unit[800]{meV}$.

In this case, we simulate the pump--probe experiment with a pump
pulse vector potential characterized by a FWHM $\tau_{\mathrm{pu}}=\unit[10]{fs}$,
amplitude $A_{0}=\unitfrac[0.5276]{V}{nm\,fs}$, and frequency $\hbar\omega_{\text{pu}}=\unit[1.55]{eV}$.
We explore the frequency spectrum within the near infrared and visible
region with the probe, revealing transitions up to a few multi-photon
resonances of the pump pulse frequency between valence and conduction
bands. The crystal orientation is the same as in Ref.~\citep{inzani2023field},
and both the pump and probe polarizations are along the direction
$\left[\begin{array}{ccc}
1 & 0 & 0\end{array}\right]$. The BZ is sampled on a $32\times32\times32$ $\boldsymbol{k}$-grid
centered at $\Gamma$, and the damping factor is set to $\lambda=\unit[0.25]{PHz}$.

\subsection{Kerr rotation: equilibrium, long-time dynamics, and damping effects}

In Fig.~\ref{fig:germanium}, we report the Kerr rotation angle,
$\theta_{K}\left(\omega\right)$, as a function of the probe frequency,
computed for our realization of weakly spin-polarized germanium. Panel
(a) shows the equilibrium Kerr rotation angle, $\theta_{K}^{\text{eq}}\left(\omega\right)$,
in the near infrared and visible frequency ranges. The complexity
of the band structure of a real material is reflected in the rich
frequency dependence of the Kerr-rotation features.

Panel (b) displays the differential Kerr rotation $\delta\theta_{K}\left(\omega,t_{\mathrm{pr}}\right)=\theta_{K}\left(\omega,t_{\mathrm{pr}}\right)-\theta_{K}^{\text{eq}}\left(\omega\right)$
as a function of the probe frequency and delay time , in the long-delay
(post-pump) regime with the undamped dynamics, i.e., with $\Upsilon=0$
. The inset shows a cut at the two-photon resonant frequency $\omega=2\omega_{\text{pu}}$,
where the maximum variation occurs. This behavior is consistent with
Ref.~\citep{inzani2023field}, where two-photon processes were found
to play the dominant role in inducing charge accumulation in the post-pump
regime. Moreover, this analysis clearly demonstrates that the Kerr
rotation angle can be used experimentally to identify the relevant
n-photon resonances for a given material. The signals expected at
the local extrema of the equilibrium Kerr rotation are barely visible,
or even invisible, on the scale of the main signal, which is much
larger in amplitude. The fast oscillations follow approximately twice
the pump frequency and originate from the interference of off-diagonal
(coherence) terms of the single-particle density matrix among $\boldsymbol{k}$
points in the valence and conduction bands at different multi-photon
resonances. Such $\boldsymbol{k}$ points host post-pump electronic
excitations. It is worth noting that, for this system, several distinct
multi-photon resonances can coexist at the same $\boldsymbol{k}$
point. The slow beating instead, follows the energy splitting induced
by the spin polarization.

Panel (c) shows the differential Kerr rotation $\delta\theta_{K}(\omega,t_{\mathrm{pr}})$
computed for the same parameters as in panel (a), except that the
damping $\Upsilon_{nm}=\delta_{nm}\tfrac{\lambda}{4}+(1-\delta_{nm})\tfrac{\lambda}{2}$
is included. In this case, the expected signal at the local extrema
of the equilibrium Kerr rotation becomes clearly visible, at least
for the most pronounced ones. Furthermore, the beating behavior disappears
because the damping acts on a shorter timescale.

\section{Conclusions and perspective }\label{sec:Conclusions}

We have developed a general and efficient theoretical framework for
computing the time-resolved magneto-optical Kerr effect in ultrafast
pump--probe experiments. Our approach, based on the Dynamical Projective
Operatorial Approach (DPOA), provides a direct link between the microscopic
non-equilibrium dynamics of a photo-excited system and its macroscopic
magneto-optical response. This formulation offers a transparent physical
picture and significant numerical advantages, being fully general
for multi-band systems with arbitrary lattice structures, spin-orbit
coupling, and symmetry breaking.

We presented a derivation of a simplified expression for the optical
conductivity that is valid after the pump pulse has become negligible,
which depends solely on the single-particle density matrix (SPDM)
at the probe pulse time. This post-pump simplification drastically
reduces the computational cost for studying long-time dynamics. Furthermore,
the framework naturally allows for the inclusion of phenomenological
damping within the SPDM dynamics, enabling the modeling of essential
relaxation processes that dominate the experimental post-pump regime.

We illustrated the applicability of this formalism by applying it
to a minimal two-band model. The numerical results successfully capture
the hallmark features of ultrafast magneto-optical Kerr effect, including
the rapid, pump-induced oscillations during the excitation and the
emergence of persistent spectral features near the pump pulse frequency
after the pulse application, such as reduced absorption due to Pauli
blocking. The calculated Kerr rotation also reveals additional structures
at the equilibrium local extrema positions, arising from the non-linear
dependence of the Kerr angle on the conductivity tensor, as well as
long-time beating patterns originating from quantum interference between
different inter-band coherences.

The application to a realization of weakly spin-polarized germanium
instead, proved the capability of the formalism to handle even the
complex band structures of real materials and that the Kerr rotation
can be used to deduct experimentally the relevant n-photon resonances
for a given specific material.

The presented framework is readily extendable to more complex materials---such
as altermagnets and topological insulators---where the interplay
of light, spin, orbital, and charge degrees of freedom leads to rich,
non-equilibrium physics. By bridging microscopic dynamics and experimental
observables, our theory provides a powerful tool for interpreting
and guiding future time-resolved magneto-optical studies across a
wide range of quantum materials.
\begin{acknowledgments}
The authors thank Jeroen van den Brink for insightful discussions.
The authors acknowledge support by MUR under Project PNRR MUR Missione
4 (SPOKE 2) TOPQIN “TOPological Qubit In driveN and reconfigurable
heterostructures”.
\end{acknowledgments}

\appendix

\section{Derivation of Eq.~\ref{eq:sigma-rho-1}}\label{sec:Derivation-sigma_rho}

In this appendix, we derive Eq.~\ref{eq:sigma-rho-1}. Setting $t_{\mathrm{fin}}=t_{\mathrm{pr}}$
in Eq.~\eqref{eq:sig_ap_1}, we obtain $\sigma_{\text{out},\text{in}}^{\left(1\right),\text{a.p.}}\left(\omega,t_{\mathrm{fin}},t_{\mathrm{pr}}\right)\rightarrow\sigma_{\text{out},\text{in}}^{\left(1\right)}\left(\omega,t_{\mathrm{pr}}\right)$
where
\begin{multline}
\sigma_{\text{out},\text{in}}^{\left(1\right)}\left(\omega,t_{\mathrm{pr}}\right)=-\frac{\mathrm{i}e}{\hbar\mathcal{V}}\times\\
\times\sum_{\boldsymbol{k}}\left\{ W_{\boldsymbol{k}}\left(\omega,t_{\mathrm{pr}}\right)+\mathrm{Tr}\left(Q_{\boldsymbol{k}}\left(\omega,t_{\mathrm{pr}}\right)\cdot S_{\boldsymbol{k}}\left(t_{\mathrm{pr}},t_{\mathrm{pr}}\right)\right)\right\} .\label{eq:sig_ap_1-1}
\end{multline}
Using Eqs.~\ref{eq:Qwt}, \ref{eq:Swt} and \ref{eq:X}, we have
\begin{multline}
\mathrm{Tr}\left(Q_{\boldsymbol{k}}\left(\omega,t_{\mathrm{pr}}\right)\cdot S_{\boldsymbol{k}}\left(t_{\mathrm{pr}},t_{\mathrm{pr}}\right)\right)=-e\mathrm{i}\times\\
\times\sum_{n,n^{\prime}}\left(\left\{ P_{\boldsymbol{k}}^{\dagger}\left(t_{\mathrm{pr}}\right)\cdot J_{\boldsymbol{k},\text{out}}^{\bar{w}}\left(\omega,t_{\mathrm{pr}}\right)\cdot P_{\boldsymbol{k}}\left(t_{\mathrm{pr}}\right)\right\} \circ\Delta F_{\boldsymbol{k}}\right)_{nn^{\prime}}\times\\
\times\left(P_{\boldsymbol{k}}^{\dagger}\left(t_{\mathrm{pr}}\right)\cdot D_{\boldsymbol{k},\text{in}}\left(t_{\mathrm{pr}}\right)\cdot P_{\boldsymbol{k}}\left(t_{\mathrm{pr}}\right)\right)_{n^{\prime}n}
\end{multline}
where, for the sake of simplicity, we have defined $J_{\boldsymbol{k},\text{out}}^{\bar{w}}\left(\omega,t_{\mathrm{pr}}\right)=J_{\boldsymbol{k},\text{out}}\left(t_{\mathrm{pr}}\right)\circ\bar{w}_{\boldsymbol{k}}\left(\omega\right)$.
Using $\varDelta F_{\boldsymbol{k},n,n^{\prime}}=f_{\boldsymbol{k},n}-f_{\boldsymbol{k},n^{\prime}}$,
we get
\begin{multline}
\mathrm{Tr}\left(Q_{\boldsymbol{k}}\left(\omega,t_{\mathrm{pr}}\right)\cdot S_{\boldsymbol{k}}\left(t_{\mathrm{pr}},t_{\mathrm{pr}}\right)\right)=-e\mathrm{i}\sum_{n,n^{\prime}}\left\{ \right.\\
\left.\left\{ P_{\boldsymbol{k}}^{\dagger}\left(t_{\mathrm{pr}}\right)\cdot J_{\boldsymbol{k},\text{out}}^{\bar{w}}\left(\omega,t_{\mathrm{pr}}\right)\cdot P_{\boldsymbol{k}}\left(t_{\mathrm{pr}}\right)\right\} _{nn^{\prime}}f_{\boldsymbol{k},n}\right.\\
\left.\left(P_{\boldsymbol{k}}^{\dagger}\left(t_{\mathrm{pr}}\right)\cdot D_{\boldsymbol{k},\text{in}}\left(t_{\mathrm{pr}}\right)\cdot P_{\boldsymbol{k}}\left(t_{\mathrm{pr}}\right)\right)_{n^{\prime}n}\right.\\
-\left.\left\{ P_{\boldsymbol{k}}^{\dagger}\left(t_{\mathrm{pr}}\right)\cdot J_{\boldsymbol{k},\text{out}}^{\bar{w}}\left(\omega,t_{\mathrm{pr}}\right)\cdot P_{\boldsymbol{k}}\left(t_{\mathrm{pr}}\right)\right\} _{nn^{\prime}}f_{\boldsymbol{k},n^{\prime}}\right.\\
\left.\left(P_{\boldsymbol{k}}^{\dagger}\left(t_{\mathrm{pr}}\right)\cdot D_{\boldsymbol{k},\text{in}}\left(t_{\mathrm{pr}}\right)\cdot P_{\boldsymbol{k}}\left(t_{\mathrm{pr}}\right)\right)_{n^{\prime}n}\right\} ,
\end{multline}
which can be written in the following compact form: 
\begin{multline}
\mathrm{Tr}\left(Q_{\boldsymbol{k}}\left(\omega,t_{\mathrm{pr}}\right)\cdot S_{\boldsymbol{k}}\left(t_{\mathrm{pr}},t_{\mathrm{pr}}\right)\right)=-e\mathrm{i}\mathrm{Tr}\left\{ \right.\\
\left.F_{\boldsymbol{k}}\cdot P_{\boldsymbol{k}}^{\dagger}\left(t_{\mathrm{pr}}\right)\cdot J_{\boldsymbol{k},\text{out}}^{\bar{w}}\left(\omega,t_{\mathrm{pr}}\right)\cdot P_{\boldsymbol{k}}\left(t_{\mathrm{pr}}\right)\cdot\right.\\
\left.P_{\boldsymbol{k}}^{\dagger}\left(t_{\mathrm{pr}}\right)\cdot D_{\boldsymbol{k},\text{in}}\left(t_{\mathrm{pr}}\right)\cdot P_{\boldsymbol{k}}\left(t_{\mathrm{pr}}\right)\right.\\
-\left.P_{\boldsymbol{k}}^{\dagger}\left(t_{\mathrm{pr}}\right)\cdot J_{\boldsymbol{k},\text{out}}^{\bar{w}}\left(\omega,t_{\mathrm{pr}}\right)\cdot P_{\boldsymbol{k}}\left(t_{\mathrm{pr}}\right)\cdot F_{\boldsymbol{k}}\cdot\right.\\
\left.P_{\boldsymbol{k}}^{\dagger}\left(t_{\mathrm{pr}}\right)\cdot D_{\boldsymbol{k},\text{in}}\left(t_{\mathrm{pr}}\right)\cdot P_{\boldsymbol{k}}\left(t_{\mathrm{pr}}\right)\right\} .
\end{multline}
Using the cyclic property of trace, Eq.~\ref{eq:rho-P} and $P_{\boldsymbol{k}}^{\dagger}\left(t\right)\cdot P_{\boldsymbol{k}}\left(t\right)=\boldsymbol{1}$,
we obtain
\begin{multline}
\mathrm{Tr}\left(Q_{\boldsymbol{k}}\left(\omega,t_{\mathrm{pr}}\right)\cdot S_{\boldsymbol{k}}\left(t_{\mathrm{pr}},t_{\mathrm{pr}}\right)\right)=\\
e\mathrm{i}\mathrm{Tr}\left\{ \left[D_{\boldsymbol{k},\text{in}}\left(t_{\mathrm{pr}}\right),J_{\boldsymbol{k},\text{out}}^{\bar{w}}\left(\omega,t_{\mathrm{pr}}\right)\right]\cdot\rho_{\boldsymbol{k}}\left(t_{\mathrm{pr}}\right)\right\} .
\end{multline}
Substituting back in Eq.~\ref{eq:sig_ap_1-1}, one obtains Eq.~\ref{eq:sigma-rho-1},
given in the main text.

It is worth noting that unlike Eq.~\ref{eq:sigma-rho-1}, Eq.~\ref{eq:sigma-rho-2}
does not need a detailed derivation, as by setting $t_{\mathrm{fin}}=t_{\mathrm{pr}}$
in Eq.~\ref{eq:sig_ap_2}, one immediately obtains it.

\section{Computing Polar Kerr Rotation Angle from the Dielectric Tensor}\label{sec:Computing-Polar-Kerr}

Let us consider a monochromatic electromagnetic plane wave impinging
perpendicularly on the surface, lying in the $xy$ plane, of a neutral
(zero free charge density) dielectric (zero free current density)
optically-linear sample in the polar Kerr geometry (magneto-optic
medium with magnetization along axis $z$) with a possible non-magnetic
optical anisotropy only in the $xy$ plane
\begin{align}
 & \mathbf{E}_{\omega}^{a}\left(\mathbf{r},t\right)=\mathbf{E}_{0}^{a}\left(\omega\right)\mathrm{e}^{\mathrm{i}\frac{\omega}{c}\left(\mathbf{n}_{a}\left(\omega\right)\cdot\mathbf{r}-ct\right)}\\
 & \mathbf{E}_{0}^{a}\left(\omega\right)=\left(E_{0x}^{a}\left(\omega\right),E_{0y}^{a}\left(\omega\right),E_{0z}^{a}\left(\omega\right)\right)\\
 & \mathbf{n}_{a}\left(\omega\right)=\left(0,0,n_{a}\left(\omega\right)\right)
\end{align}
where $\mathbf{E}_{\omega}^{a}\left(\mathbf{r},t\right)$ is the electric
field, $a=i,\:i',\:r$ stands for incident, reflected and refracted
wave, respectively, and $n_{a}\left(\omega\right)=-1\:(i),\:1\:(i'),\:-n\left(\omega\right)\:(r)$
is the (generally complex) refractive index of the traversed medium,
respectively. $\mathbf{n}_{a}\left(\omega\right)$ has only the $z$
component for all three waves because of the kinematic conservation
of the momentum parallel to the interface (Snell law for flat, homogeneous,
and static interface in linear optics) \citep{Jackson1998,BornWolf1999,Zangwill2012,Hecht2017}.
Normal incidence gives $E_{0z}^{i}\left(\omega\right)=0$.

The relevant Maxwell's equations read as
\begin{align}
 & \nabla\cdot\mathbf{D}_{\omega}^{a}\left(\mathbf{r},t\right)=0\Rightarrow\mathbf{n}_{a}\left(\omega\right)\cdot\mathbf{D}_{0}^{a}\left(\omega\right)=0\label{eq:div_D}\\
 & \nabla\times\mathbf{E}_{\omega}^{a}\left(\mathbf{r},t\right)=-\frac{\partial\mathbf{B}_{\omega}^{a}}{\partial t}\left(\mathbf{r},t\right)\label{eq:curl_E}\\
 & \nabla\times\mathbf{B}_{\omega}^{a}\left(\mathbf{r},t\right)=\mu_{0}\frac{\partial\mathbf{D}_{\omega}^{a}}{\partial t}\left(\mathbf{r},t\right)\label{eq:curl_B}
\end{align}
where $\mathbf{D}_{\omega}^{a}\left(\mathbf{r},t\right)=\mathbf{D}_{0}^{a}\left(\omega\right)\mathrm{e}^{\mathrm{i}\frac{\omega}{c}\left(\mathbf{n}_{a}\left(\omega\right)\cdot\mathbf{r}-ct\right)}$
is the electric displacement field and $\mathbf{B}_{\omega}\left(\mathbf{r},t\right)=\mathbf{B}_{0}\left(\omega\right)\mathrm{e}^{\mathrm{i}\frac{\omega}{c}\left(\mathbf{n}\left(\omega\right)\cdot\mathbf{r}-ct\right)}$
is the magnetic induction field. At finite frequencies, we can use
$\mu_{0}$ instead of the permeability tensor $\boldsymbol{\mu}$
\citep{Landau:1960aa,Pershan:1967aa,Dolgov:1989aa,Oppeneer:2001aa}.

The polar Kerr geometry, and a possible non-magnetic optical anisotropy
only in the $xy$ plane, require that the dielectric tensor of the
sample (magneto-optic medium), $\boldsymbol{\epsilon}\left(\omega\right)$,
describing the relation between the electric and electric displacement
fields, $\mathbf{D}_{0}^{a}\left(\omega\right)=\varepsilon_{0}\boldsymbol{\epsilon}\left(\omega\right)\cdot\mathbf{E}_{0}\left(\omega\right)$,
has this generic form
\begin{equation}
\boldsymbol{\epsilon}\left(\omega\right)=\begin{pmatrix}\epsilon_{xx}\left(\omega\right) & \epsilon_{xy}\left(\omega\right) & 0\\
\epsilon_{yx}\left(\omega\right) & \epsilon_{yy}\left(\omega\right) & 0\\
0 & 0 & \epsilon_{zz}\left(\omega\right)
\end{pmatrix}.
\end{equation}
Then, Eq.~(\ref{eq:div_D}) implies that $E_{0z}^{i'}\left(\omega\right)=E_{0z}^{r}\left(\omega\right)=0$.

Applying the curl ($\nabla\times$) to Eq.~(\ref{eq:curl_E}) and
using Eq.~(\ref{eq:curl_B}), one can straightforwardly show that
\begin{equation}
\left(n^{2}\left(\omega\right)\boldsymbol{1}-\boldsymbol{\epsilon}\left(\omega\right)\right)\cdot\mathbf{E}_{0}^{r}\left(\omega\right)=0.\label{eq:n2E}
\end{equation}
 Being Eq. \ref{eq:n2E} valid for any $\mathbf{E}_{0}^{i}\left(\omega\right)$
and, consequently, for any $\mathbf{E}_{0}^{r}\left(\omega\right)$,
it requires that
\begin{equation}
\det\left(n^{2}\left(\omega\right)\boldsymbol{1}-\boldsymbol{\epsilon}\left(\omega\right)\right)=0,
\end{equation}
which gives the characteristics of the two principal propagation modes
in the sample
\begin{align}
 & n_{\pm}\left(\omega\right)=n_{0}\left(\omega\right)\sqrt{1\pm\gamma\left(\omega\right)\sqrt{\alpha+\eta^{2}\left(\omega\right)+\gamma_{od}\left(\omega\right)}},\\
 & \underline{E}_{0y}^{r}\left(\omega\right)=\zeta_{\pm}\left(\omega\right)\underline{E}_{0x}^{r}\left(\omega\right),\\
 & \zeta_{\pm}\left(\omega\right)=-\eta\left(\omega\right)\pm\sqrt{\alpha+\eta^{2}\left(\omega\right)+\gamma_{od}\left(\omega\right)},
\end{align}
where we have rewritten the components of the dielectric tensor in
the following way
\begin{align}
 & \epsilon_{xx,yy}\left(\omega\right)=\epsilon_{d}\left(\omega\right)\pm\delta\epsilon_{d}\left(\omega\right),\\
 & \epsilon_{xy}\left(\omega\right)=\epsilon_{od}\left(\omega\right),\\
 & \epsilon_{yx}\left(\omega\right)=\alpha\epsilon_{od}\left(\omega\right)+\delta\epsilon_{od}\left(\omega\right),\:\alpha=\pm1,
\end{align}
and used the following definitions
\begin{align}
 & n_{0}\left(\omega\right)=\sqrt{\epsilon_{d}\left(\omega\right)},\\
 & \gamma\left(\omega\right)=\frac{\epsilon_{od}\left(\omega\right)}{\epsilon_{d}\left(\omega\right)},\\
 & \gamma_{od}\left(\omega\right)=\frac{\delta\epsilon_{od}\left(\omega\right)}{\epsilon_{od}\left(\omega\right)},\\
 & \eta\left(\omega\right)=\frac{\delta\epsilon_{d}\left(\omega\right)}{\epsilon_{od}\left(\omega\right)}.
\end{align}
These definitions allow to properly take into account (i) the two
main conditions, linear birefringence ($\alpha=+1$, reciprocal anisotropy)
and circular birefringence ($\alpha=-1$, gyrotropy), and (ii) the
actual relative relevance of the terms in the great majority of real
 experimental conditions where (for a given $\alpha$) we have
\begin{align}
 & \left|\delta\epsilon_{d}\left(\omega\right)\right|\ll\left|\epsilon_{d}\left(\omega\right)\right|\Rightarrow\left|\gamma_{d}\left(\omega\right)\right|\ll1,\\
 & \left.\begin{array}{l}
\left|\epsilon_{od}\left(\omega\right)\right|\ll\left|\epsilon_{d}\left(\omega\right)\right|\\
\left|\delta\epsilon_{od}\left(\omega\right)\right|\ll\left|\epsilon_{od}\left(\omega\right)\right|
\end{array}\right\} \Rightarrow\left|\gamma_{od}\left(\omega\right)\right|\ll\left|\gamma_{d}\left(\omega\right)\right|\ll1.
\end{align}
Accordingly, we have
\begin{multline}
n_{\pm}\left(\omega\right)\cong n_{0}\left(\omega\right)\\
\pm\frac{1}{2}\sqrt{\alpha}\gamma\left(\omega\right)n_{0}\left(\omega\right)\sqrt{{\scriptstyle 1+\alpha\eta^{2}\left(\omega\right)}}\left(1+\frac{1}{2}\alpha\frac{\gamma_{od}\left(\omega\right)}{{\scriptstyle 1+\alpha\eta^{2}\left(\omega\right)}}\right),
\end{multline}
and
\begin{multline}
\zeta_{\pm}\left(\omega\right)\cong-\eta\left(\omega\right)\\
\pm\sqrt{\alpha}\sqrt{{\scriptstyle 1+\alpha\eta^{2}\left(\omega\right)}}\left(1+\frac{1}{2}\alpha\frac{\gamma_{od}\left(\omega\right)}{{\scriptstyle 1+\alpha\eta^{2}\left(\omega\right)}}\right).
\end{multline}
The approximate formulas take into account that by a proper rotation
of the coordinate system around the $z$ axis (of an angle $\theta=\frac{1}{2}\arctan\left(\frac{2\eta\left(\omega\right)}{1+\alpha+\gamma_{od}\left(\omega\right)}\right)$
with respect to the $x$ axis), we can reduce $\delta\epsilon_{d}\left(\omega\right)$,
and, accordingly, also $\eta\left(\omega\right)$, to zero, moving
the anisotropy of the diagonal elements to a symmetric component of
the off-diagonal elements. The expressions of $\zeta_{\pm}\left(\omega\right)$
 clarifies why the sign of $\alpha$ determines the linearity ($\alpha=+1\Rightarrow\zeta_{\pm}\left(\omega\right)\:\textrm{predominatly real}$)
or the circularity ($\alpha=-1\Rightarrow\zeta_{\pm}\left(\omega\right)\:\textrm{predominantly imaginary}$)
of the birefringence.

Now, any incident wave can be formally decomposed in the two principal
propagation modes in the sample
\begin{align}
 & \mathbf{E}_{0}^{i}\left(\omega\right)=\left(1-w\left(\omega\right)\right)\mathbf{E}_{0}^{i+}\left(\omega\right)+w\left(\omega\right)\mathbf{E}_{0}^{i-}\left(\omega\right),\\
 & \mathbf{E}_{0}^{i+}\left(\omega\right)=\left(1,\zeta_{+}\left(\omega\right)\right)E_{0x}^{i}\left(\omega\right),\\
 & \mathbf{E}_{0}^{i-}\left(\omega\right)=\left(1,\zeta_{-}\left(\omega\right)\right)E_{0x}^{i}\left(\omega\right),\\
 & w\left(\omega\right)=\frac{\zeta\left(\omega\right)-\zeta_{+}\left(\omega\right)}{\zeta_{-}\left(\omega\right)-\zeta_{+}\left(\omega\right)},\\
 & \zeta\left(\omega\right)=\frac{E_{0y}^{i}\left(\omega\right)}{E_{0x}^{i}\left(\omega\right)},
\end{align}
that allows us to use the Fresnel relations for normal incidence \citep{Jackson1998,BornWolf1999,Zangwill2012,Hecht2017}
to easily compute the reflected wave as the two principal propagation
modes will be reflected with the following coefficients
\begin{align}
 & \mathbf{E}_{0}^{i'\pm}\left(\omega\right)=r_{\pm}\left(\omega\right)\mathbf{E}_{0}^{i\pm}\left(\omega\right),\\
 & r_{\pm}\left(\omega\right)=\frac{1-n_{\pm}\left(\omega\right)}{1+n_{\pm}\left(\omega\right)},
\end{align}
that, to the first order in $\gamma\left(\omega\right)$ and $\gamma_{od}\left(\omega\right)$,
assume the following expressions 
\begin{multline}
r_{\pm}\left(\omega\right)\cong r_{0}\left(\omega\right)\\
\mp\sqrt{\alpha}\gamma\left(\omega\right)\frac{n_{0}\left(\omega\right)}{\left(1+n_{0}\left(\omega\right)\right)^{2}}\sqrt{{\scriptstyle 1+\alpha\eta^{2}\left(\omega\right)}}\left(1+\frac{1}{2}\alpha\frac{\gamma_{od}\left(\omega\right)}{{\scriptstyle 1+\alpha\eta^{2}\left(\omega\right)}}\right),
\end{multline}
where $r_{0}\left(\omega\right)=\frac{1-n_{0}\left(\omega\right)}{1+n_{0}\left(\omega\right)}$.
Accordingly, the reflected wave reads as
\begin{align}
 & E_{0x}^{i'}\left(\omega\right)=\left(\left(1-w\left(\omega\right)\right)r_{+}\left(\omega\right)+w\left(\omega\right)r_{-}\left(\omega\right)\right)E_{0x}^{i}\left(\omega\right),\nonumber \\
 & E_{0y}^{i'}\left(\omega\right)=\left(1-w\left(\omega\right)\right)r_{+}\left(\omega\right)\zeta_{+}\left(\omega\right)E_{0x}^{i}\left(\omega\right)\\
 & +w\left(\omega\right)r_{-}\left(\omega\right)\zeta_{-}\left(\omega\right)E_{0x}^{i}\left(\omega\right),
\end{align}
and we have, to the first order in $\gamma_{od}\left(\omega\right)$,
\begin{align}
 & E_{0x}^{i'}\left(\omega\right)=r_{0}\left(\omega\right)E_{0x}^{i}\left(\omega\right)\\
 & +\gamma\left(\omega\right)\frac{n_{0}\left(\omega\right)}{\left(1+n_{0}\left(\omega\right)\right)^{2}}\left(\zeta\left(\omega\right)-\eta\left(\omega\right)\right)E_{0x}^{i}\left(\omega\right),\\
 & E_{0y}^{i'}\left(\omega\right)=\zeta\left(\omega\right)r_{0}\left(\omega\right)E_{0x}^{i}\left(\omega\right)\\
 & +\gamma\left(\omega\right)\eta\left(\omega\right)\left(\zeta\left(\omega\right)-\eta\left(\omega\right)\right)\frac{n_{0}\left(\omega\right)}{\left(1+n_{0}\left(\omega\right)\right)^{2}}E_{0x}^{i}\left(\omega\right)\\
 & +\gamma\left(\omega\right)\frac{n_{0}\left(\omega\right)}{\left(1+n_{0}\left(\omega\right)\right)^{2}}\left(\alpha+\eta^{2}\left(\omega\right)+\gamma_{od}\left(\omega\right)\right)E_{0x}^{i}\left(\omega\right),
\end{align}
that defines the following Jones (reflection) matrix
\begin{align}
 & r_{xx}\left(\omega\right)=r_{0}\left(\omega\right)\left(1-\eta\left(\omega\right)\gamma\left(\omega\right)\frac{n_{0}\left(\omega\right)}{1-n_{0}^{2}\left(\omega\right)}\right),\\
 & r_{xy}\left(\omega\right)=r_{0}\left(\omega\right)\gamma\left(\omega\right)\frac{n_{0}\left(\omega\right)}{1-n_{0}^{2}\left(\omega\right)},\\
 & r_{yx}\left(\omega\right)=r_{0}\left(\omega\right)\gamma\left(\omega\right)\left(\alpha+\gamma_{od}\left(\omega\right)\right)\frac{n_{0}\left(\omega\right)}{1-n_{0}^{2}\left(\omega\right)},\\
 & r_{yy}\left(\omega\right)=r_{0}\left(\omega\right)\left(1+\eta\left(\omega\right)\gamma\left(\omega\right)\frac{n_{0}\left(\omega\right)}{1-n_{0}^{2}\left(\omega\right)}\right).
\end{align}

We can now compute the Kerr rotation angle \citep{Argyres1955_FaradayKerr,Pershan:1967aa,ErskineStern1973_GdKerr,BeaurepaireMerleDaunoisBigot1996_NiDemag,ZvezdinKotov1997_ModernMagnetooptics,QiuBader2000_SMOKE,Oppeneer:2001aa,ZhangHubnerLefkidisBaiGeorge2009_TRMOKE,WuJiangWeng2010_SpinDynamics}
for a linear polarization along the $x$ axis ($E_{0y}^{i}\left(\omega\right)=0\Rightarrow\zeta\left(\omega\right)=0$)
\begin{equation}
\theta_{\mathrm{K}}^{\phi=0}\left(\omega\right)=\Re\arctan\frac{r_{yx}\left(\omega\right)}{r_{xx}\left(\omega\right)},
\end{equation}
where $\phi$ is the angle formed by the polarization with the $x$
axis. Up to the first order in $\gamma\left(\omega\right)$ and using
that $\arctan\left(x\right)\simeq x$ for $\left|x\right|\ll1$, $\theta_{\mathrm{K}}^{\phi=0}\left(\omega\right)$
read as
\begin{multline}
\theta_{\mathrm{K}}^{\phi=0}\left(\omega\right)=\Re\left(\gamma\left(\omega\right)\left(\alpha+\gamma_{od}\left(\omega\right)\right)\frac{n_{0}\left(\omega\right)}{1-n_{0}^{2}\left(\omega\right)}\right)\\
=\Re\left[\frac{\epsilon_{yx}\left(\omega\right)}{\sqrt{\epsilon_{d}\left(\omega\right)}\left(1-\epsilon_{d}\left(\omega\right)\right)}\right]
\end{multline}

For any other polarization angle $\phi$, we can obtain the Kerr rotation
angle by rotating the coordinate system in order to have the actual
polarization to coincide with the new $x$ axis and use the formulas
just derived. This would require to rotate accordingly the dielectric
tensor
\begin{align}
 & \epsilon{}_{xx}^{\phi}=\epsilon_{xx}\cos^{2}\phi+(\epsilon_{xy}+\epsilon_{yx})\sin\phi\cos\phi+\epsilon_{yy}\sin^{2}\phi,\\
 & \epsilon{}_{xy}^{\phi}=\epsilon_{xy}\cos^{2}\phi-\epsilon_{yx}\sin^{2}\phi+(\epsilon_{yy}-\epsilon_{xx})\sin\phi\cos\phi,\\
 & \epsilon{}_{yx}^{\phi}=\epsilon_{yx}\cos^{2}\phi-\epsilon_{xy}\sin^{2}\phi+(\epsilon_{yy}-\epsilon_{xx})\sin\phi\cos\phi,\\
 & \epsilon{}_{yy}^{\phi}=\epsilon_{xx}\sin^{2}\phi-(\epsilon_{xy}+\epsilon_{yx})\sin\phi\cos\phi+\epsilon_{yy}\cos^{2}\phi.
\end{align}
that gives
\begin{equation}
\theta_{\mathrm{K}}^{\phi}\left(\omega\right)=\Re\left[\frac{\epsilon_{yx}^{\phi}\left(\omega\right)}{\sqrt{\epsilon_{d}\left(\omega\right)}\left(1-\epsilon_{d}\left(\omega\right)\right)}\right],
\end{equation}
where $\epsilon_{d}\left(\omega\right)$, being half of the trace
of the dielectric tensor, is an invariant under such a rotation.

According to this, the Kerr rotation angle is finite, with opposite
signs, in both standard cases: for a mainly ($\left|\delta\epsilon_{od}\left(\omega\right)\right|\ll\left|\epsilon_{od}\left(\omega\right)\right|$)
linear birefringent ($\alpha=+1$) sample and for a mainly circular
birefringent ($\alpha=-1$) sample. On the other hand, in the case
of pure linear birefringence ($\alpha=+1$, $\left|\delta\epsilon_{od}\left(\omega\right)\right|=0$,
reciprocal anisotropy: $\epsilon_{yx}\left(\omega\right)=\epsilon_{xy}\left(\omega\right)$),
the Kerr rotation angle can be reduced to zero by choosing appropriately
the impinging linear polarization, i.e., along the $x$ axis of the
coordinate system in which the dielectric tensor of the sample can
be made diagonal. While, in the case of mainly circular birefringence
($\alpha=-1$, $\left|\delta\epsilon_{od}\left(\omega\right)\right|\ll\left|\epsilon_{od}\left(\omega\right)\right|$,
quasi gyrotropy: $\epsilon_{yx}\left(\omega\right)\cong-\epsilon_{xy}\left(\omega\right)$),
the Kerr rotation angle cannot be reduced to zero by any choice of
the linear polarization. This occurrence makes the latter case, quasi
gyrotropy, much more relevant and interesting as, in this case, the
Kerr rotation angle (i) becomes a fundamental measure of the gyrotropic
magneto-optical response of the system, beyond standard linear birefringence,
and (ii) reveals to be an invariant for certain magnetic semiconductors
(usual ingredients are a finite spin-orbit coupling and broken time-reversal
symmetry) in the same line as the Chern number is for topological
insulators.

Now, while theoretically, to check the (magneto-optical) gyrotropy
of a sample is enough to compute the degree of antisymmetric anisotropy
through the definition of the polar (magneto-optical) Kerr rotation
angle
\begin{equation}
\theta_{\mathrm{K}}\left(\omega\right)=\Re\left[\frac{\epsilon_{yx}\left(\omega\right)-\epsilon_{xy}\left(\omega\right)}{2\sqrt{\epsilon_{d}\left(\omega\right)}\left(1-\epsilon_{d}\left(\omega\right)\right)}\right],\label{eq:Kerr_app_eps}
\end{equation}
experimentally, we need to average over all possible linear polarizations
to integrate out the linear birefringence/symmetric anisotropy component
and single out the circular birefringence/antisymmetric anisotropy
component \citep{weber2024all}. Such an averaging procedure results
in
\begin{equation}
\bar{\theta}_{\mathrm{K}}\left(\omega\right)=\frac{1}{2\pi}\int_{0}^{2\pi}\theta_{\mathrm{K}}^{\phi}\left(\omega\right)d\phi=\theta_{\mathrm{K}}\left(\omega\right),
\end{equation}
which in turn justifies the definition of the polar Kerr rotation
angle given in Eq.~\ref{eq:Kerr_app_eps}.

\bibliographystyle{apsrev4-2}
\bibliography{biblio}

\end{document}